\documentclass{PoS}

\let\ifpdf\relax

\usepackage{graphicx}
\usepackage{amsmath}
\usepackage{amsfonts}
\usepackage{subfigure}
\usepackage{wrapfig}
\usepackage{epsfig}
\usepackage{afterpage,float}
\usepackage{cite}
\usepackage{ifpdf} 

\newcommand{\be}{\begin{equation}}
\newcommand{\ee}{\end{equation}}
\newcommand{\bea}{\begin{eqnarray}}
\newcommand{\eea}{\end{eqnarray}}

\def\half{{\textstyle{1\over2}}}

\def\bar{\overline}

\def\tilde{\widetilde}

\def\half{{\scriptstyle \raise.15ex\hbox{${1\over2}$}}}

\newcommand{\beq}{\begin{equation}}
\newcommand{\eeq}{\end{equation}}

\newcommand{\real}{\relax{\rm I\kern-.18em R}}

\title{Twelve fundamental and two sextet fermion flavors
 }

\ShortTitle{Twelve fundamental and two sextet fermion flavors}

\author{Zolt\'an Fodor\\
        Department of Physics, University of Wuppertal\\
        Gau$\beta$strasse 20, D-42119, Germany\\
        J\"ulich Supercomputing Center, Forschungszentrum\\
        J\"ulich, D-52425 J\"ulich, Germany\\
        Email: \email{fodor@bodri.elte.hu}}

\author{Kieran Holland\\
        Department of Physics, University of the Pacific\\
        3601 Pacific Ave, Stockton CA 95211, USA\\
        Email: \email{kholland@pacific.edu}}

\author{Julius Kuti*
\hphantom{\speaker{J.~Kuti and C.~Schroeder}}\\
        Department of Physics 0319, University of California, San Diego\\
        9500 Gilman Drive, La Jolla, CA 92093, USA\\
        E-mail: \email{jkuti@ucsd.edu}}

\author{D\'aniel N\'ogr\'adi\\
        Institute for Theoretical Physics, E\"otv\"os University\\
        H-1117 Budapest, Hungary\\
        Email: \email{nogradi@bodri.elte.hu}}

\author{Chris Schroeder*\\
        Physical Sciences Directorate, Lawrence Livermore National Laboratory\\
        Livermore, California 94550, USA\\
        E-mail: \email{chris.schroeder@gmail.com}}
        
\author{Chik Him Wong\\
        Department of Physics 0319, University of California, San Diego\\
        9500 Gilman Drive, La Jolla, CA 92093, USA\\
        E-mail: \email{rickywong@physics.ucsd.edu} }

\abstract{
We report extended simulation results and their new analysis in
two important gauge theories with twelve fermion flavors  
in the fundamental SU(3) color representation and two fermions in the sextet representation. 
We probe the $N_f=12$ model with respect to the conformal window using mass deformed 
finite size scaling (FSS) theory driven by  the fermion mass anomalous dimension. 
Our  results at fixed gauge coupling
show problems with the conformal scenario of the $N_f=12$ model. 
In the sextet model with two flavors, under the conformal hypothesis, we determine 
large values for the anomalous fermion mass dimension with  $\gamma\geq 1$. Since our sextet analysis
favors the chiral symmetry breaking hypothesis without conformality, 
the large exponent $\gamma$ could play an important role in understanding the composite Higgs mechanism.
The new results discussed here include
our extended data sets and exceed what was presented at the conference.
\vskip 0.005 in
}

\FullConference{The XXIX International Symposium on Lattice Field Theory - Lattice 2011\\
July 10-16, 2011,
Squaw Valley, Lake Tahoe, California}

\begin{document}

\section{Introduction}

New physics at the Large Hadron Collider could be discovered in the form of some new strongly-interacting gauge
theory with a composite Higgs mechanism, an idea which was outside experimental 
reach when it was first introduced as an attractive scenario beyond the Standard Model
~\cite{Weinberg:1979bn,Susskind:1978ms,Dimopoulos:1979es,
Eichten:1979ah,Farhi:1980xs,Holdom:1984sk,Yamawaki:1985zg,Appelquist:1987fc,Miransky:1996pd}. 
The original framework has  been expanded by new
explorations of the multi-dimensional theory space of nearly conformal gauge theories
~\cite{Caswell:1974gg,Banks:1981nn,Appelquist:2003hn, Sannino:2004qp,
Dietrich:2005jn,Luty:2004ye,Dietrich:2006cm,Kurachi:2006ej,Mojaza:2010cm}
where systematic and non-perturbative lattice studies 
play a very important role.
Interesting models require the theory to be very
close to, but below, the conformal window, with the gauge coupling
slowly evolving over a large energy range. 
The non-perturbative knowledge of the critical number of flavors $N_f^{crit}$, separating the conformal phase from the phase of the composite
Higgs mechanism with chiral symmetry breaking ($\chi{\rm SB}$), is essential and 
this has generated much interest with many old and new lattice
studies~\cite{Fodor:2009wk,Fodor:2011tw,Fodor:2011tu,Appelquist:2007hu,Appelquist:2009ty,
Appelquist:2011dp,Appelquist:2009ka,Deuzeman:2008sc,
Deuzeman:2009mh,Deuzeman:2011pa,Hasenfratz:2009ea,Hasenfratz:2010fi,Cheng:2011ic,Jin:2009mc,Jin:2010vm,
Aoki:2012kr,Catterall:2007yx,Catterall:2008qk,Hietanen:2008mr,
Hietanen:2009az,DelDebbio:2010hx,Bursa:2010xn,DelDebbio:2010ze,DelDebbio:2011kp, 
Shamir:2008pb,DeGrand:2010na,DeGrand:2011cu,DeGrand:2012yq,Kogut:2010cz,Sinclair:2010be, Bilgici:2009kh,
Itou:2010we,Yamada:2009nt,Hayakawa:2010yn,Gavai:1985wi,Attig:1987mf,Kogut:1987ai,
Meyer:1990xd,Damgaard:1997ut,Kim:1992pk,Brown:1992fz,Iwasaki:2003de}.  
The position of the conformal window with respect to the much discussed model of twelve fermions in the fundamental representation 
remains controversial with recent efforts 
from several lattice groups~\cite{Fodor:2009wk,Fodor:2011tw,Fodor:2011tu,
Appelquist:2007hu,Appelquist:2009ty,Appelquist:2011dp, 
Appelquist:2009ka, Deuzeman:2008sc, Deuzeman:2009mh,Deuzeman:2011pa, 
Hasenfratz:2009ea, Hasenfratz:2010fi,Cheng:2011ic,Jin:2009mc,Jin:2010vm,Aoki:2012kr}.
The position of the $N_f=2$ sextet model with respect to the conformal window
also remains unsettled~\cite{Fodor:2011tw,DeGrand:2010na,DeGrand:2011cu,DeGrand:2012yq,Kogut:2010cz,Sinclair:2010be}. 

Probing the conformal and $\chi{\rm SB}$ hypotheses we use two different 
strategies to deal with finite volume dependence. 
The first strategy extrapolates the spectrum to infinite volume at fixed fermion mass $m$ where the leading 
finite size correction  is exponentially small and determined by the lowest mass which has pion quantum numbers.
From the mass spectrum of the infinite volume extrapolation we can probe 
the mass deformed conformal scaling behavior and compare with $\chi SB$ behavior when the fermion mass is varied 
in the infinite volume limit. 
The second strategy takes full advantage of the conformal
FSS behavior without intrinsic scale when pressing against the $m=0$ critical surface at fixed finite size $L$. 
Different from the first strategy, the finite volume corrections are not exponentially small and a much larger 
data set is analyzed closer to the critical surface.
This is used in the $N_f=12$ model, significantly extending our previously reported results~\cite{Fodor:2011tu}.
We will also briefly summarize our main results in the sextet model where the first strategy is sufficient, 
since only runs at the lowest fermion mass show consistent and detectable finite volume dependence. 

We have used the tree-level Symanzik-improved gauge action for all simulations in this paper.
The conventional $\beta=6/g^2$ lattice gauge coupling is defined as the overall
factor in front of the well-known terms of the Symanzik lattice action.  Its value is $\beta=2.2$ for all simulations
reported here for the $N_f=12$ model. In the sextet model results are reported at $\beta=3.2$.
The link variables in the staggered fermion matrix were exponentially smeared with  two
stout steps~\cite{Morningstar:2003gk}; the precise definition of the staggered stout action was given in~\cite{Aoki:2005vt}.  
The RHMC and HMC algorithms were deployed in all runs.
For molecular dynamics time evolution we applied multiple time scales~\cite{Urbach:2005ji} and the
Omelyan integrator~\cite{Takaishi:2005tz}.
Our error analysis of  hadron masses used correlated fitting with double jackknife procedure on the covariance matrices.
The time histories of the fermion condensate, the plaquette, 
and correlators are used to monitor autocorrelation times in the simulations.

\section{Twelve fermion flavors in fundamental  SU(3) color representation}
Extending our earlier work~\cite{Fodor:2011tu}, we have new simulation results at $\beta=2.2$ in the 
fermion mass range ${\rm m=0.002-0.025}$ at lattice volumes
$20^3\times 40$,  $24^3\times 48$,  $28^3\times 56$, $40^3\times 80$, and $48^3\times 96$. 
The extended data base now spans the  ${\rm m=0.002-0.035}$  range.
The  new lowest fermion mass runs at  ${\rm m=0.002,0.004,0.006,0.008}$ can be used in the conformal FSS analysis
which over the full set would correspond to a variation of the pion correlation length in the 2.5 to 20 range in the infinite volume limit.
Results from the two lowest masses at ${\rm m=0.002,0.004}$ are not included in the current analysis and will be reported later.
For further control on finite volume dependence,  large $48^3\times 96$ 
runs were continued to two thousand trajectories at ${\rm m=0.01}$ and ${\rm m=0.015}$.  
Four runs were 
further added at $40^3\times 80$ with ${\rm m=0.01, 0.15, 0.02, 0.025}$.
The new and refreshed data set  was subjected to conformal FSS analysis and $\chi{\rm SB}$ tests of the
$\langle \bar{\psi}\psi\rangle$ chiral condensate.

\subsection{The phase diagram in the ${\bf\rm\beta-m}$ plane}
The phase structure of the model remains controversial, particularly the critically important weak coupling phase.
In addition to our spectroscopy and conformal FSS runs, we ran extensive scans at various fixed volumes and fixed fermion masses
to explore the bulk phase structure. The bare coupling $\beta$ was varied over a large range starting 
from very small $\beta$ values deep in the strong
coupling regime to the weak coupling phase at $\beta=2.2$ where the conformal and $\chi{\rm SB}$ analyses
were done.
Fermion masses ${\rm m=0.007,0.01, 0.02}$ were used
in the scans with spatial lattice sizes $L=8,12,16,20,24,32$ running a large densely spaced set in the important
and much discussed intermediate region in transit from strong coupling to weak coupling.
These scans were also extended to $N_f=2,4,6,8,10,12,14,16$ flavors. 
We will briefly summarize next what is known about the bulk lattice phase
structure.
\begin{figure}[hbt!]
\begin{center}
\begin{tabular}{cc}
\includegraphics[width=0.39\textwidth]{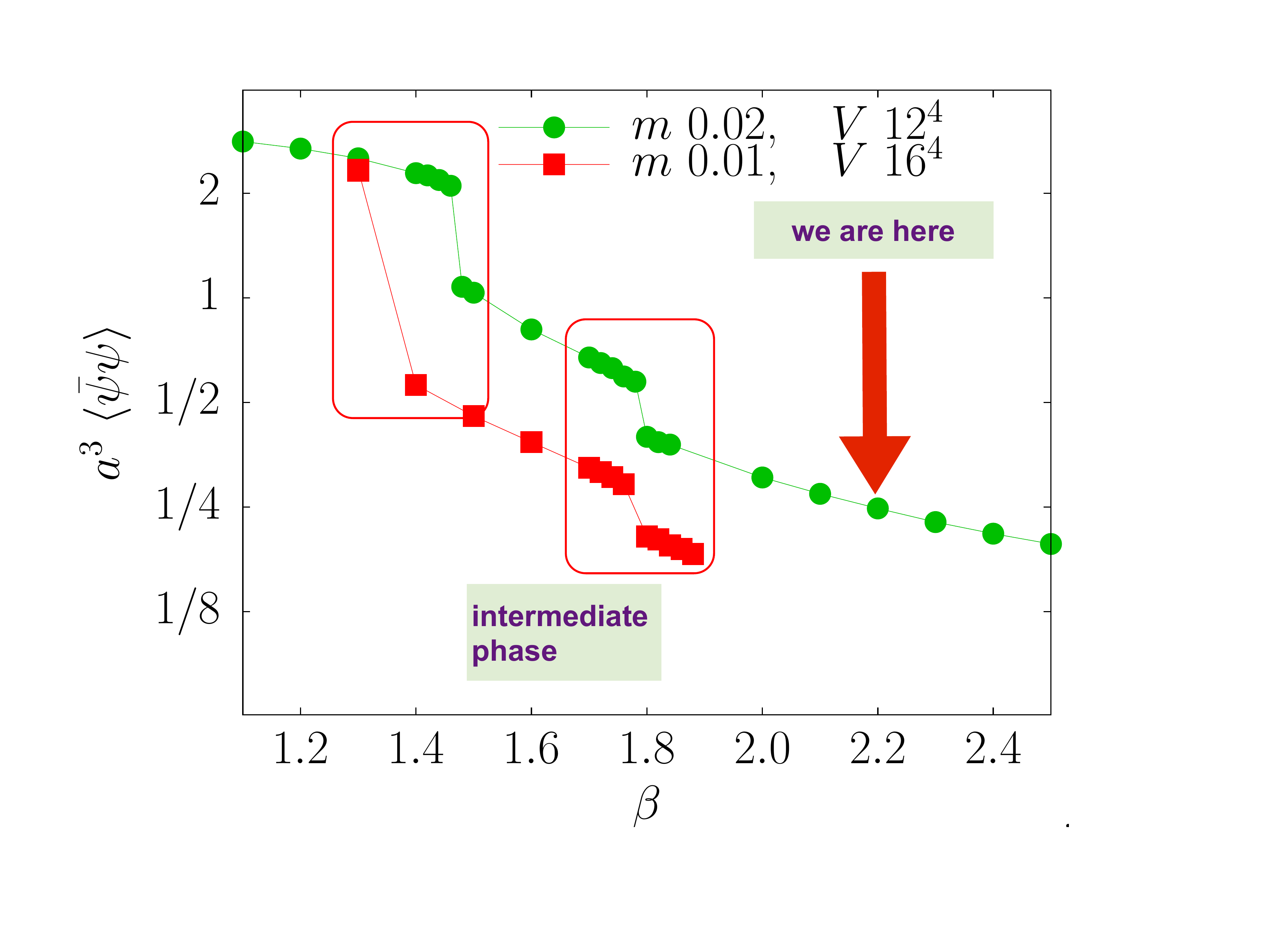}&
 \includegraphics[width=0.35\textwidth]{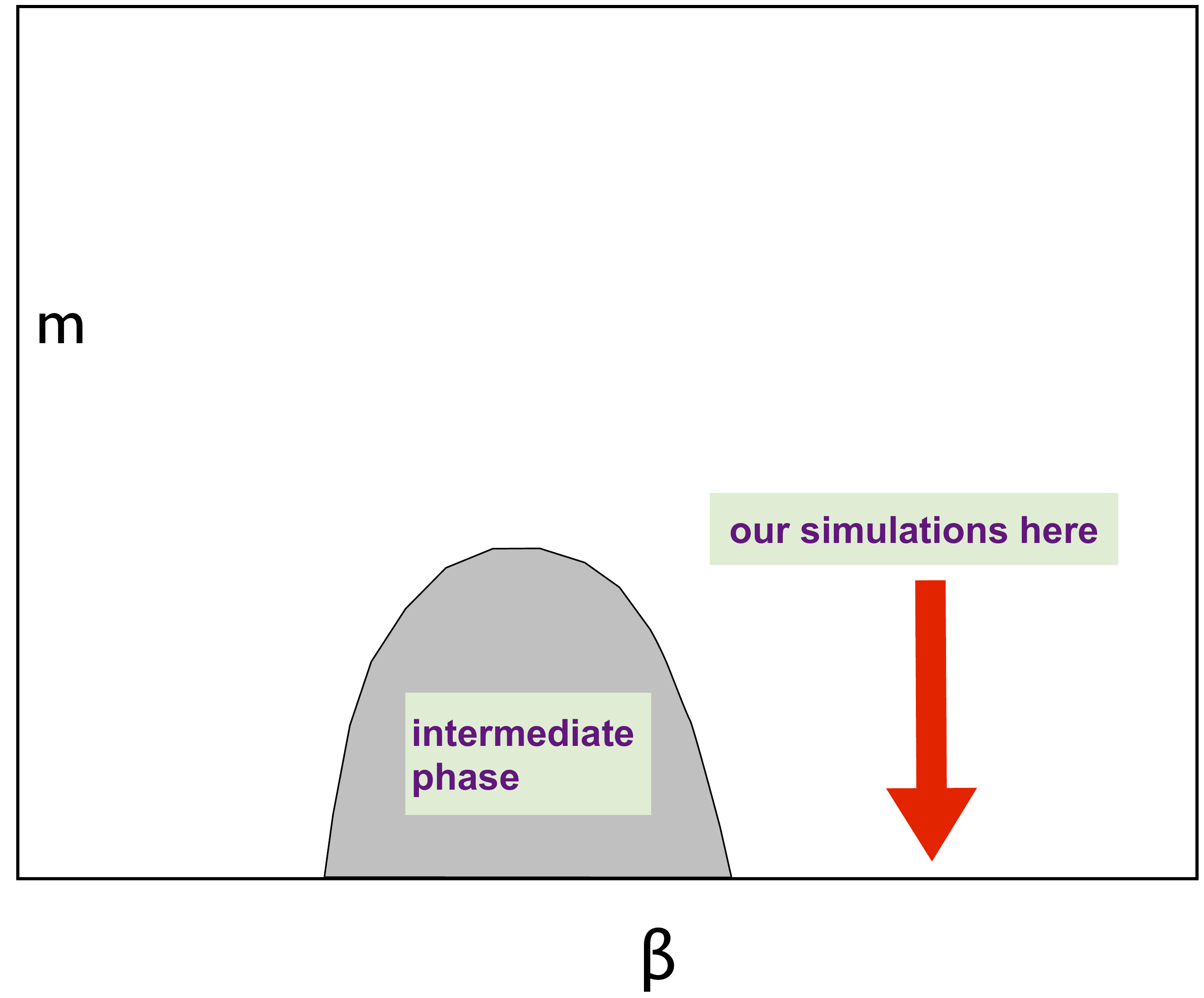}
\end{tabular}
\end{center}
\vskip -0.2in
\caption{\footnotesize On the left, scans of the phase diagram by monitoring the chiral condensate are plotted as a function of  $\beta$ 
at two different fermion masses. The schematic bulk phase diagram is sketched on the right.}
\label{fig:PhaseDiagram1}
\end{figure}

Two representative scans of the bulk behavior of the chiral condensate 
$\langle \bar{\psi}\psi\rangle$  are shown  in Figure~\ref{fig:PhaseDiagram1} as we vary $\beta$ from strong to weak coupling.
Three distinct regions emerge at fixed volume and fixed fermion mass showing strong coupling behavior 
for $\beta < 1.4$ with a large chiral condensate, an intermediate phase for
$1.4 < \beta <1.8$ with sudden drop in $\langle \bar{\psi}\psi\rangle$, and a weak coupling phase for $\beta > 1.8$ with 
further drop in $\langle \bar{\psi}\psi\rangle$. A similar structure of three regimes was also seen in scans at $N_f=8$. Our 
physics simulations were done well inside the weak coupling phase at $\beta=2.2$ as indicated in Figure~\ref{fig:PhaseDiagram1}. 
A similar structure has been observed
independently by Deuzemen et al.~\cite{Deuzeman:2011pa} and Cheng et al.~\cite{Cheng:2011ic}. 
The newfound 
order parameter of broken shift symmetry in the intermediate phase is the most interesting development 
in the study of the esoteric intermediate phase~\cite{Cheng:2011ic}. It only exists in a finite interval of the lattice gauge coupling for
small enough fermion masses, as schematically sketched in Figure~\ref{fig:PhaseDiagram1}. 
The real interest is, of course, in the nature of the weak coupling phase. 
Based on axial U(1) symmetry considerations, arguments were presented in~\cite{Deuzeman:2011pa} 
in favor of conformal symmetry in the 
weak coupling phase. This argument was criticized and refuted  in~\cite{Cheng:2011ic} based on new details of the
broken shift symmetry with chiral symmetry restoration they discovered
 at zero temperature in the bulk intermediate phase.

Cheng et al. also presented their  first weak coupling results on the Polyakov loop, the chiral condensate,
and spectroscopy as indications of conformal symmetry in the weak coupling phase.
The blocked Polyakov was reported to jump from zero to a large finite value in crossing to the weak 
coupling phase ~\cite{Cheng:2011ic}.
A confining potential was found  in the intermediate phase with broken shift symmetry which turned into Coulomb potential 
without string tension in the weak coupling phase~\cite{Cheng:2011ic}.
It was also noted that the observed chiral condensate and the related Dirac spectrum show the recovery of exact chiral 
symmetry in the massless fermion limit of the weak coupling phase consistent with observed degeneracies of parity partners
even at finite fermion masses.

The results in~\cite{Cheng:2011ic} suggesting a chirally symmetric deconfined conformal phase 
are in contradiction with what we reported 
earlier~\cite{Fodor:2011tu} and further confirmed in the 
extended new analysis. Using lattice volumes several times larger
we find a vanishing Polyakov loop at zero temperature in the weak coupling phase and a confining 
potential at a pion mass which is lower than in~\cite{Cheng:2011ic}. We also find the parity partners split 
at finite fermion mass. Our findings in large volumes are consistent with a chirally broken weak coupling phase.
As a first step to resolve the contradictions, large finite volume effects acknowledged in~\cite{Cheng:2011ic} 
will have to be brought under better control.

In the next sub-sections we will briefly summarize our results on the chiral condensate, 
the finite temperature phase transition and  tests of
the conformal hypothesis in the weak coupling phase.

\subsection{Chiral condensate and $\chi{\rm SB}$ test}

\begin{figure}[t!]
\begin{center}
\begin{tabular}{ccc}
\includegraphics[height=4.2cm]{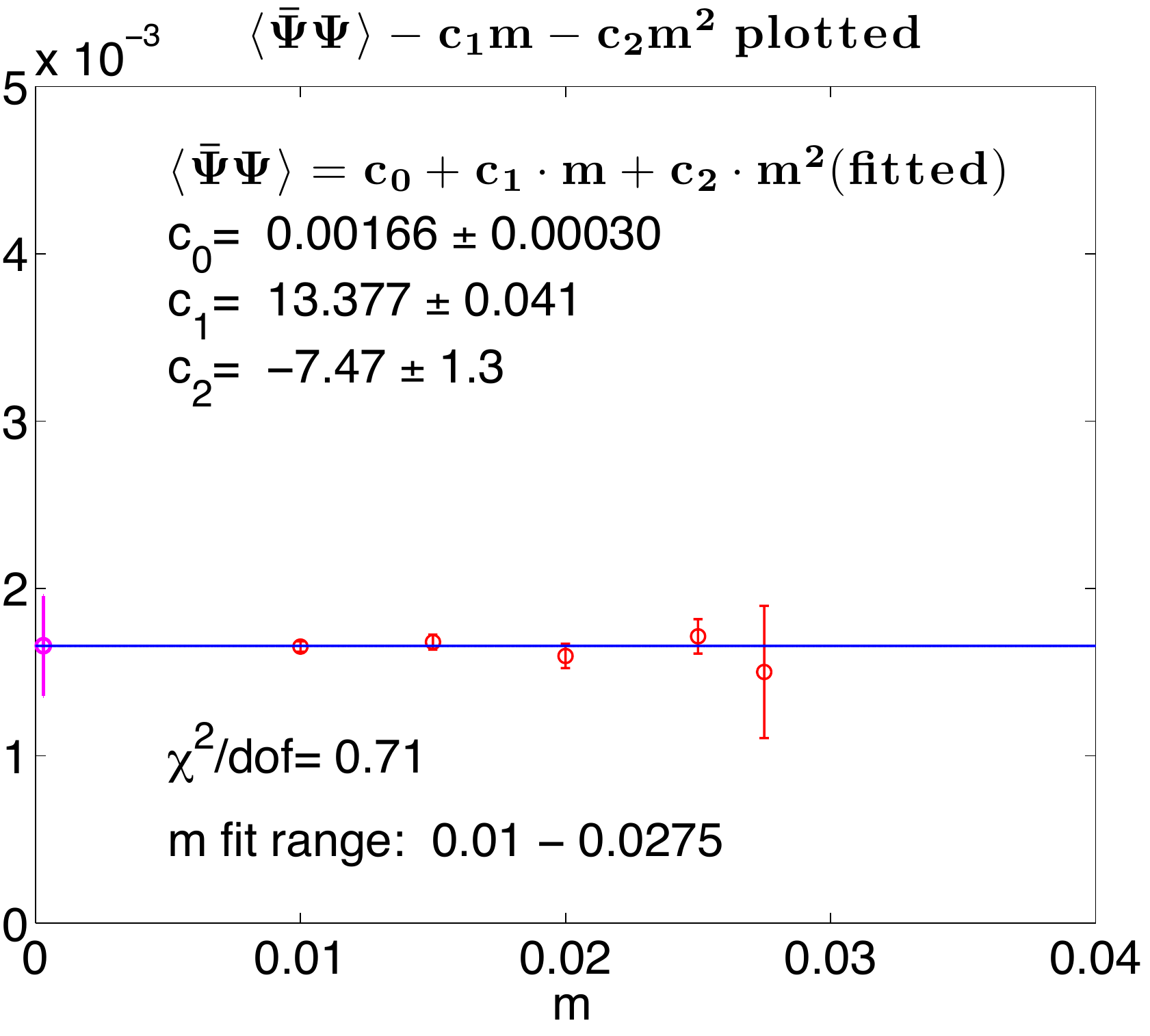}&
\includegraphics[height=4.2cm]{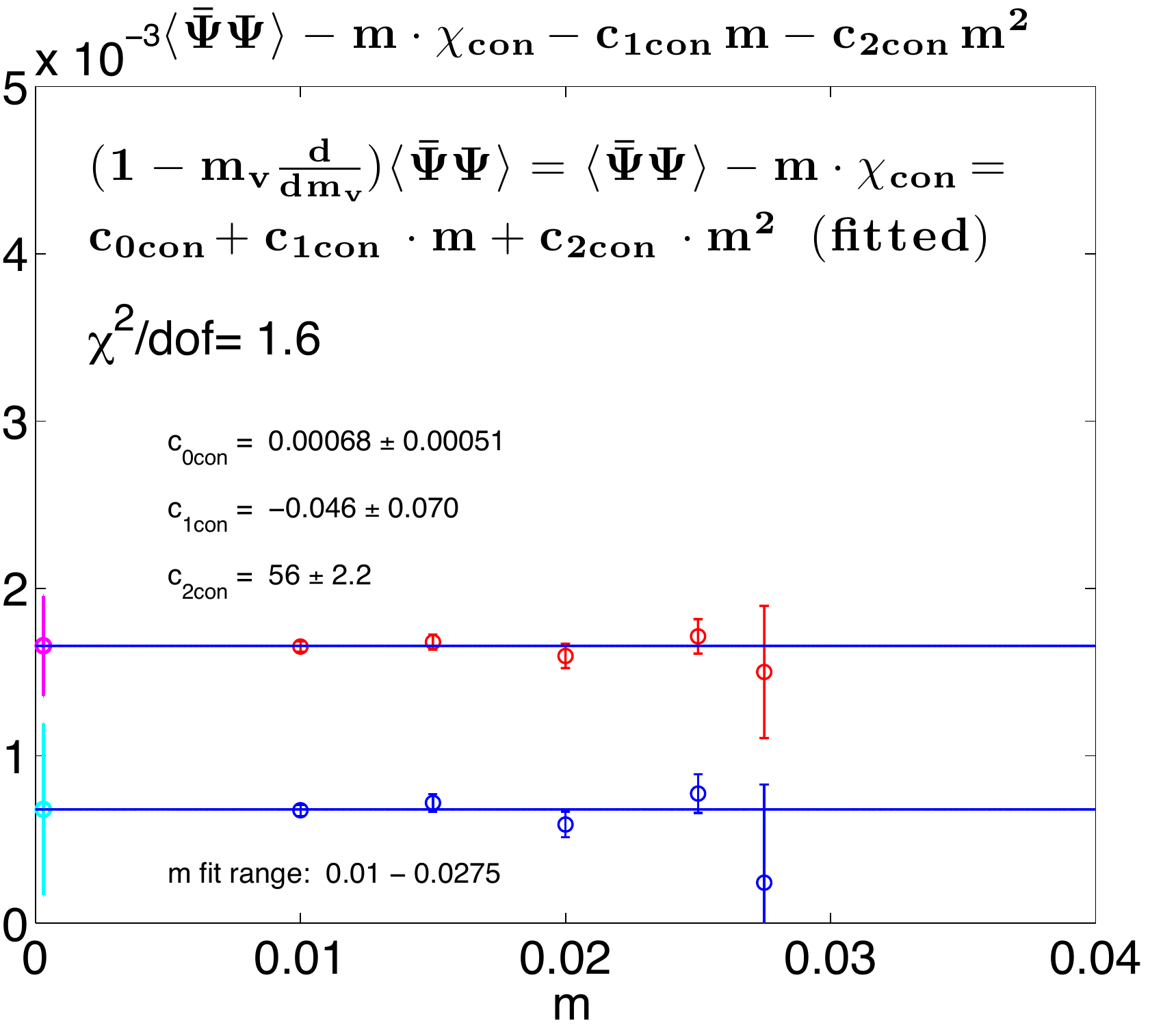}&
\includegraphics[height=4.2cm]{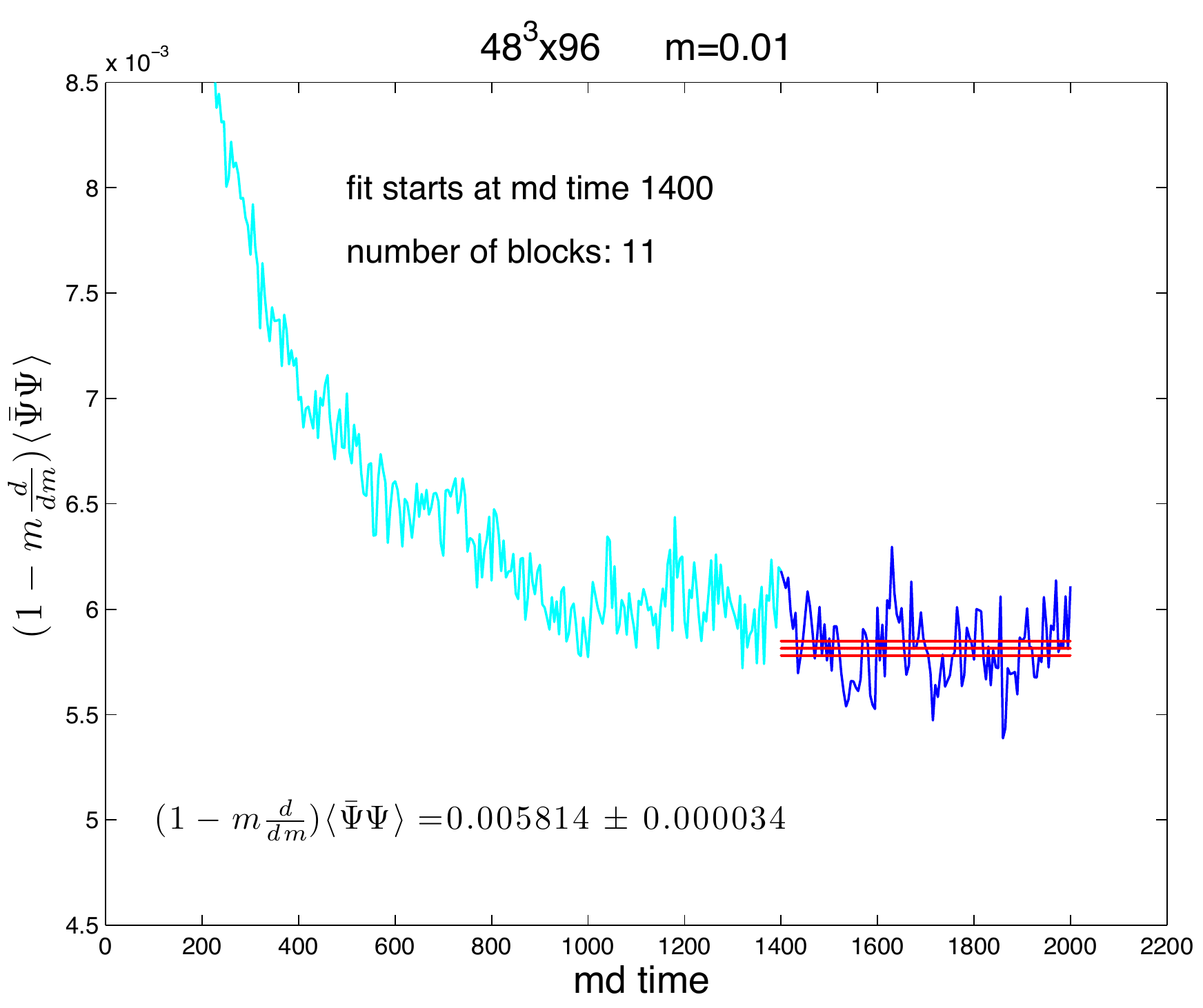}
\end{tabular}
\end{center}
\caption{{
\footnotesize 
The second order polynomial fit to the chiral condensate is shown on the left plot in subtracted 
form as explained in~\cite{Fodor:2011tu}. 
The middle plot is the quadratic fit to $\langle \bar{\psi}\psi\rangle - m\cdot\chi_{con}$
directly measured from zero momentum sum rules and independently from functions of the inverse staggered fermion
matrix. The right side plot shows the thermal history of the subtracted form of the condensate at the lowest fermion 
mass on the largest lattice.
}}
\label{fig:PbPNf12}
\end{figure}

In the extended new analysis the chiral condensate (Fig.~\ref{fig:PbPNf12})  remains consistent with $\chi{\rm SB}$ in the massless fermion limit. 
Small changes in the fits are mostly driven by the two lowest fermion masses at ${\rm m=0.01}$ and 
${\rm m=0.015}$  where runs on the largest $48^3\times 96$ lattices were extended and new runs
at higher masses on $40^3\times 80$ lattices were added. A slight drift at the lowest $m=0.01$  fermion mass 
was detected in the connected part of the condensate even
after 1,400 trajectories (i.e., MD time units) and the run was continued to 2,000 trajectories. 
Since the finite volume analysis is incomplete,  the largest volumes are used in the fitting range of the fermion masses.
Finite volume extrapolations have to be completed before definitive
conclusions can be drawn to establish a  non-vanishing condensate  $\langle\bar\psi\psi\rangle$ in the 
$m \rightarrow 0$ limit.

Details of the fitting procedure and the notation in Figure~\ref{fig:PbPNf12} were explained earlier~\cite{Fodor:2011tu}.
The chiral condensate has a spectral representation~\cite{Banks:1979yr}
where the UV-divergent integral is written in a twice-subtracted form~\cite{Leutwyler:1992yt}.
The UV contribution, which is divergent when the cutoff  $\mu ~ a^{-1}$ is removed, has a linear term  $\approx a^{-2}\cdot m$  and 
there is a third-order term  $\approx\! m^3$ which is small and hard to detect for small $m$. 
The IR finite contributions to the chiral Lagrangian
have a constant term $\approx\! B F^2$, a linear term $\approx\! B^2\cdot m$, a quadratic term $\approx B^3F^{-2}\cdot m^2$, and higher 
order terms, in addition to logarithmic corrections generated from chiral loops~\cite{Bijnens:2009qm}. 
We kept a constant IR term and the linear and second order terms with UV and IR contributions. 
The second order fit in  Figure~\ref{fig:PbPNf12}
gives a non-vanishing condensate in the chiral limit which is roughly consistent with 
the GMOR~\cite{GellMann:1968rz} relation  $\langle\bar{\psi}\psi\rangle=2F^2B$ 
with the measured low value of $F$  and the value of $B$ from logarithmic fit to the Goldstone pion. 
The deficit between the two sides of the GMOR relation is sensitive to the fitting procedure 
and uncertainties in the determination of  $B$. Trying to identify chiral logs is 
beyond the scope of our simulation range. 
For independent determination, we studied the subtracted chiral condensate operator defined with the help
of the connected part $\chi_{con}$ of the chiral susceptibility $\chi$ as defined in ~\cite{Fodor:2011tu}.
The removal of the derivative term significantly reduces the 
dominant linear part of the $\langle \bar{\psi}\psi\rangle$ condensate. 
Although the two independent determinations give consistent non-vanishing results in the chiral limit,  we 
cannot consider the results definitive. The drop of the chiral limit intercepts after extended runs is
noted in comparison with earlier results ~\cite{Fodor:2011tu}.

\subsection{Finite temperature transition}
We present some preliminary results from our extended studies of the finite temperature
transition. If the ground state of the model has ${\rm \chi SB}$, a phase transition is expected
at some finite temperature to restore the chiral symmetry in the limit of massless fermions. 
\begin{figure}[ht]
\begin{center}
\begin{tabular}{cc}
\includegraphics[width=0.4\textwidth]{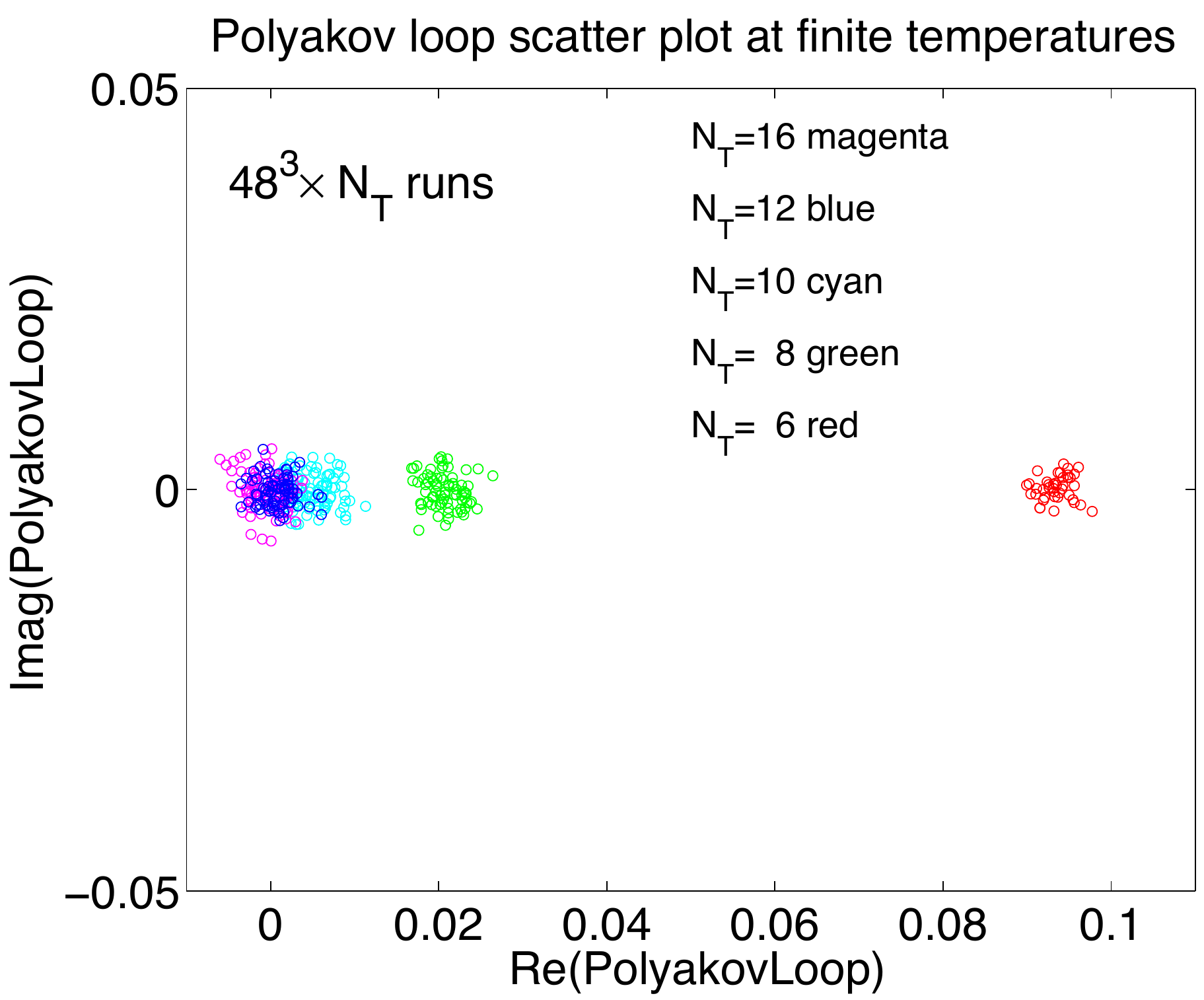}&
\includegraphics[width=0.451\textwidth]{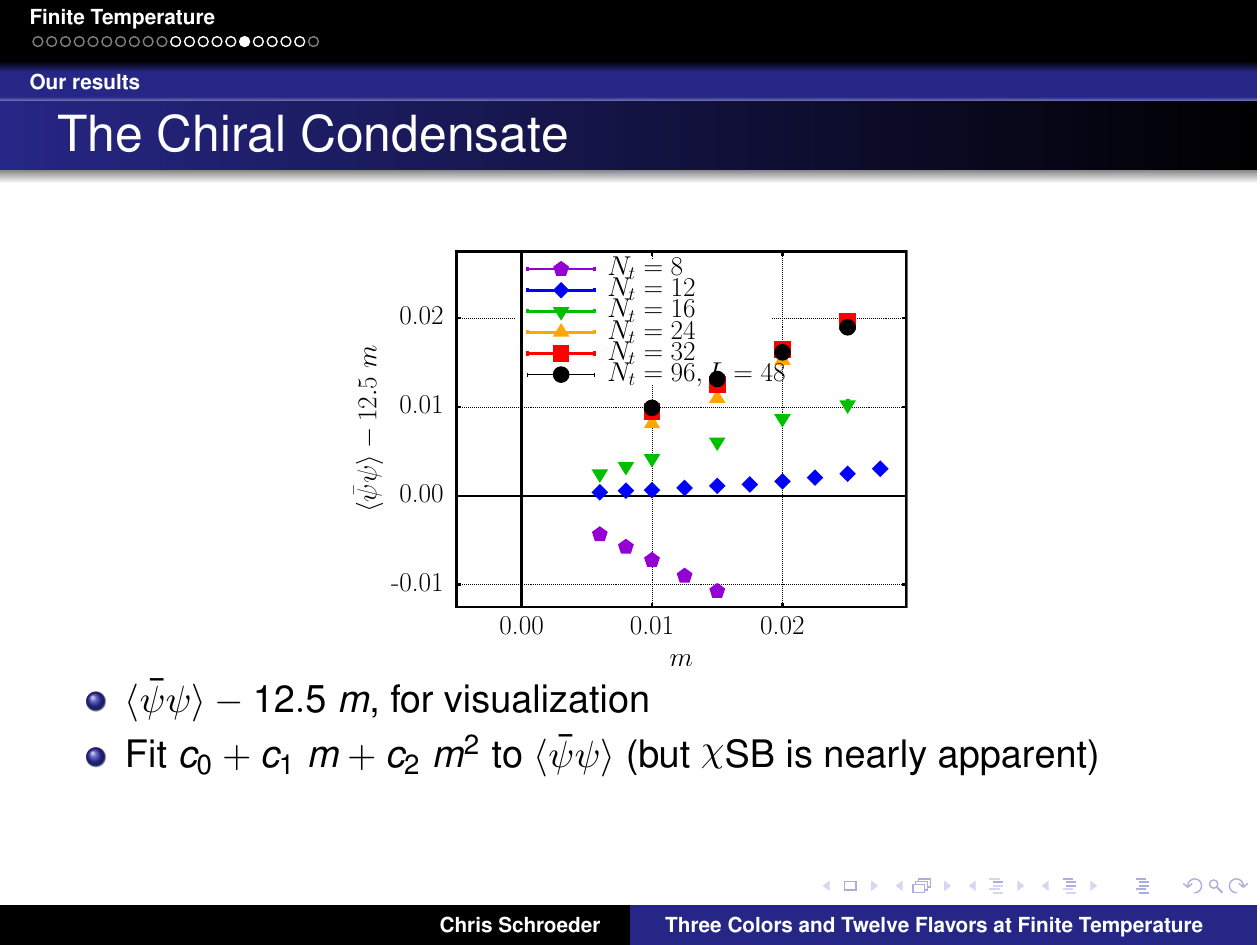}
\end{tabular}
\end{center}
\caption{{\footnotesize The scatter plot on the left side shows the Polyakov loop in the time direction as the temperature is
varied using  a sequence of lattice sizes $48^3\times N_T$ at fixed $m=0.01$  and $\beta=2.2$ with $N_T$ 
varied in the $N_T=6,8,10,12,16$ range. On the right side plot the subtracted chiral condensate is plotted at 
 $\beta=2.2$ for a sequence of lattice sizes $48^3\times N_T$ as the fermion mass is varied.
 The subtracted linear term $12.5\cdot m$, chosen for the presentation of the data, is not determined from fits.}}
\label{fig:ploop}
\end{figure} 
Based on universality arguments~\cite{Pisarski:1983ms}  the transition would be expected  to be of first order. 
This is not entirely clear and warrants further investigations.
On our largest $48^3\times N_T$ lattices, at fixed $m=0.01$ and $\beta=2.2$, the temperature was varied through an $N_T$ sequence 
while the scatter plot of the Polyakov
loop was monitored along the euclidean time (inverse temperature) direction in each run. The chiral condensate $\langle \bar{\psi}\psi\rangle$
was also monitored in the runs. As the temperature is increased a clear sudden transition is observed in the $N_T=6-10$ region
where  the Polyakov loop
distribution jumps from the origin to a scatter plot with non-vanishing real part. 
It would be difficult to reconcile this jump, as shown in Figure~\ref{fig:ploop}, with conformal
behavior in the zero temperature bulk phase. 

Although we have results for temperature scans at multiple gauge couplings, 
fermion masses, and spatial volumes, all consistent with a finite temperature transition,
caution is necessary before firm conclusions can be reached. 
Confirming the existence of the ${\rm \chi SB}$ phase transition will require the $m\rightarrow 0$  limits 
of  $\langle \bar{\psi}\psi\rangle$ and the Polyakov loop distribution.
The chiral condensate is a good order parameter for the transition. The Polyakov loop, like in QCD, could detect deconfinement
in the transition with unsettled interpretation as order parameter.
The behavior of the renormalized Polyakov loop is consistent with the scatter plot of  Figure~\ref{fig:ploop}.

\subsection{Conformal finite size scaling analysis}
\begin{figure}[!t]
\begin{center}
\begin{tabular}{cc}
\includegraphics[height=5.5cm]{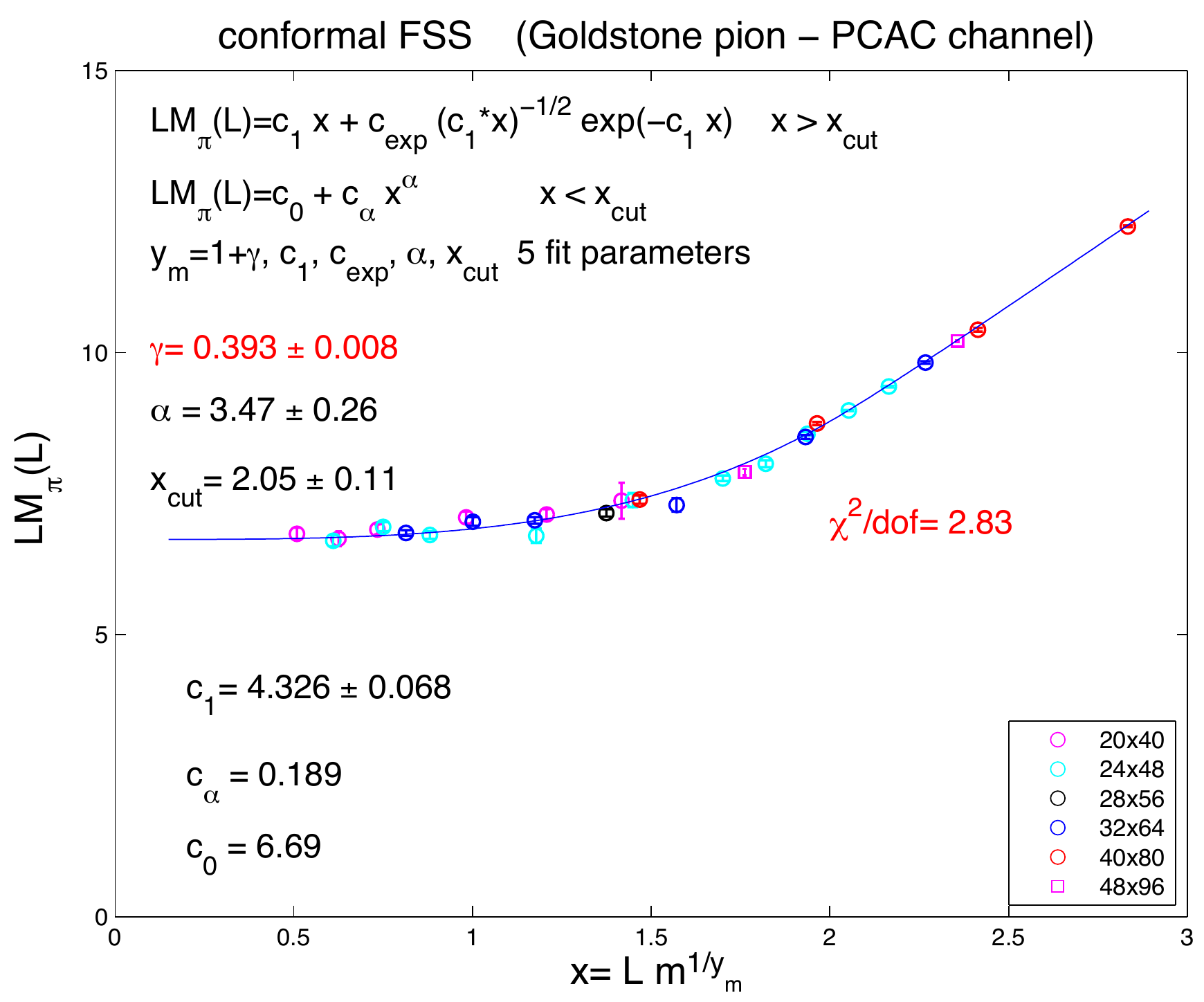}&
\includegraphics[height=5.5cm]{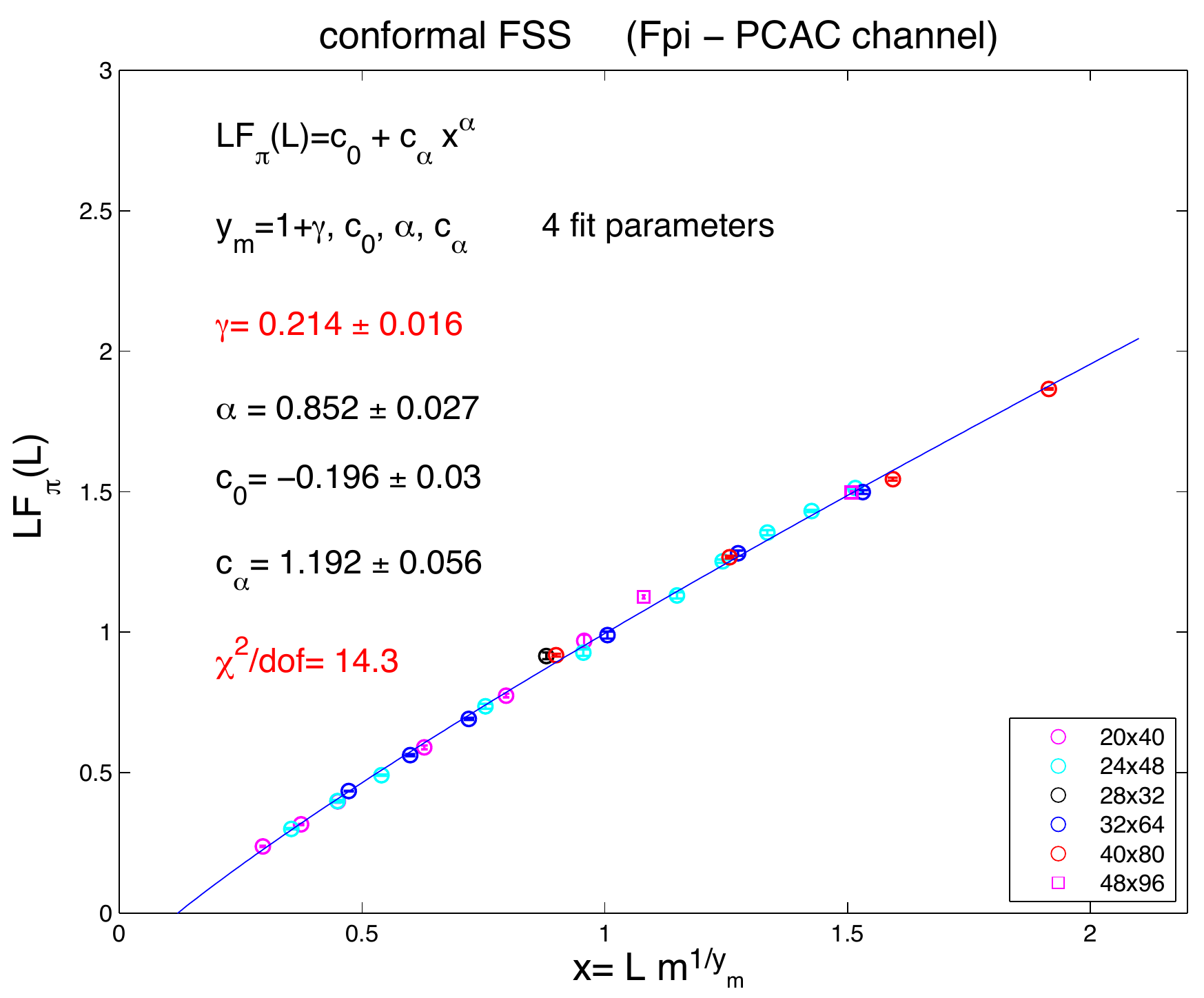}\\
\includegraphics[height=5.5cm]{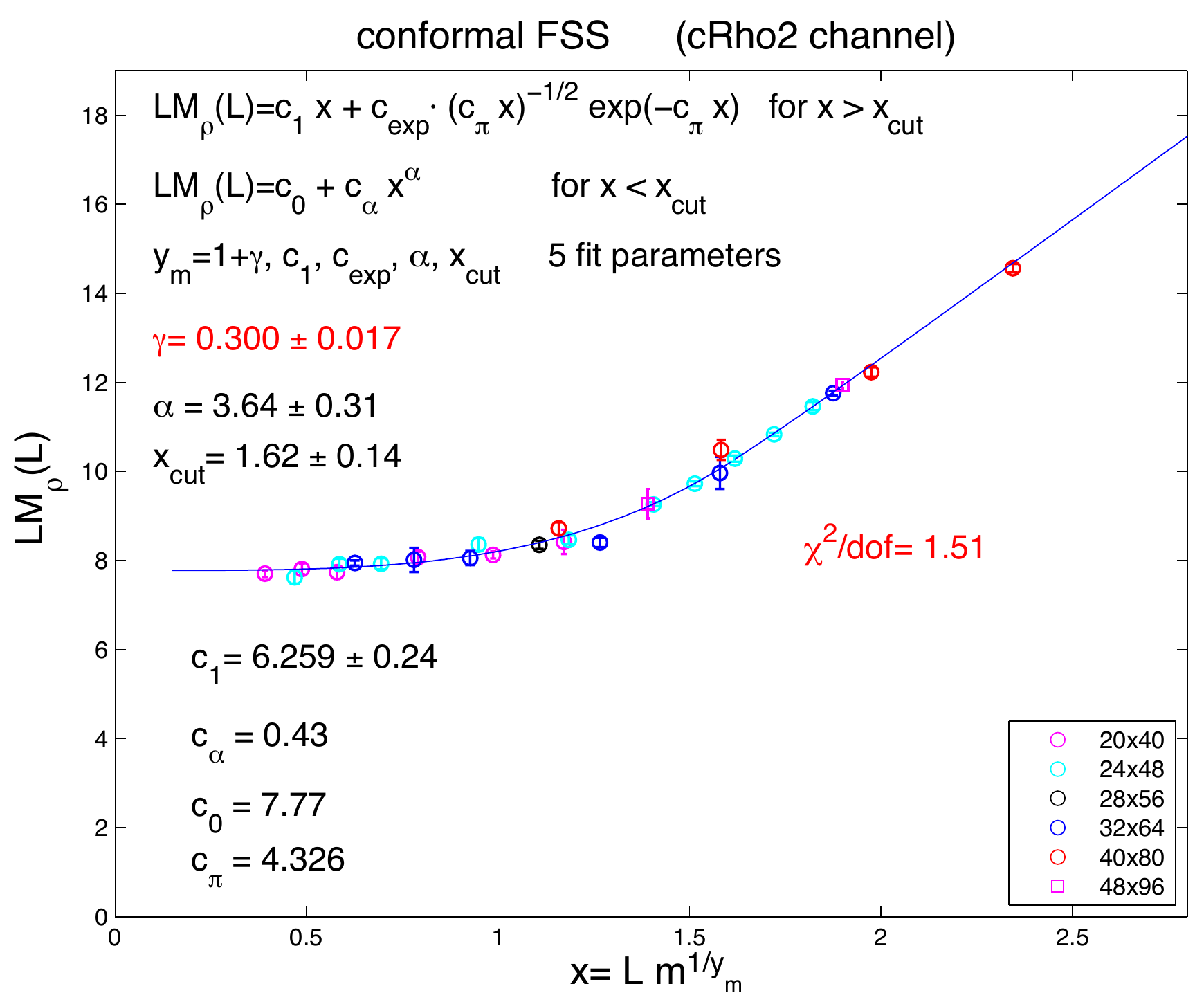}&
\includegraphics[height=5.5cm]{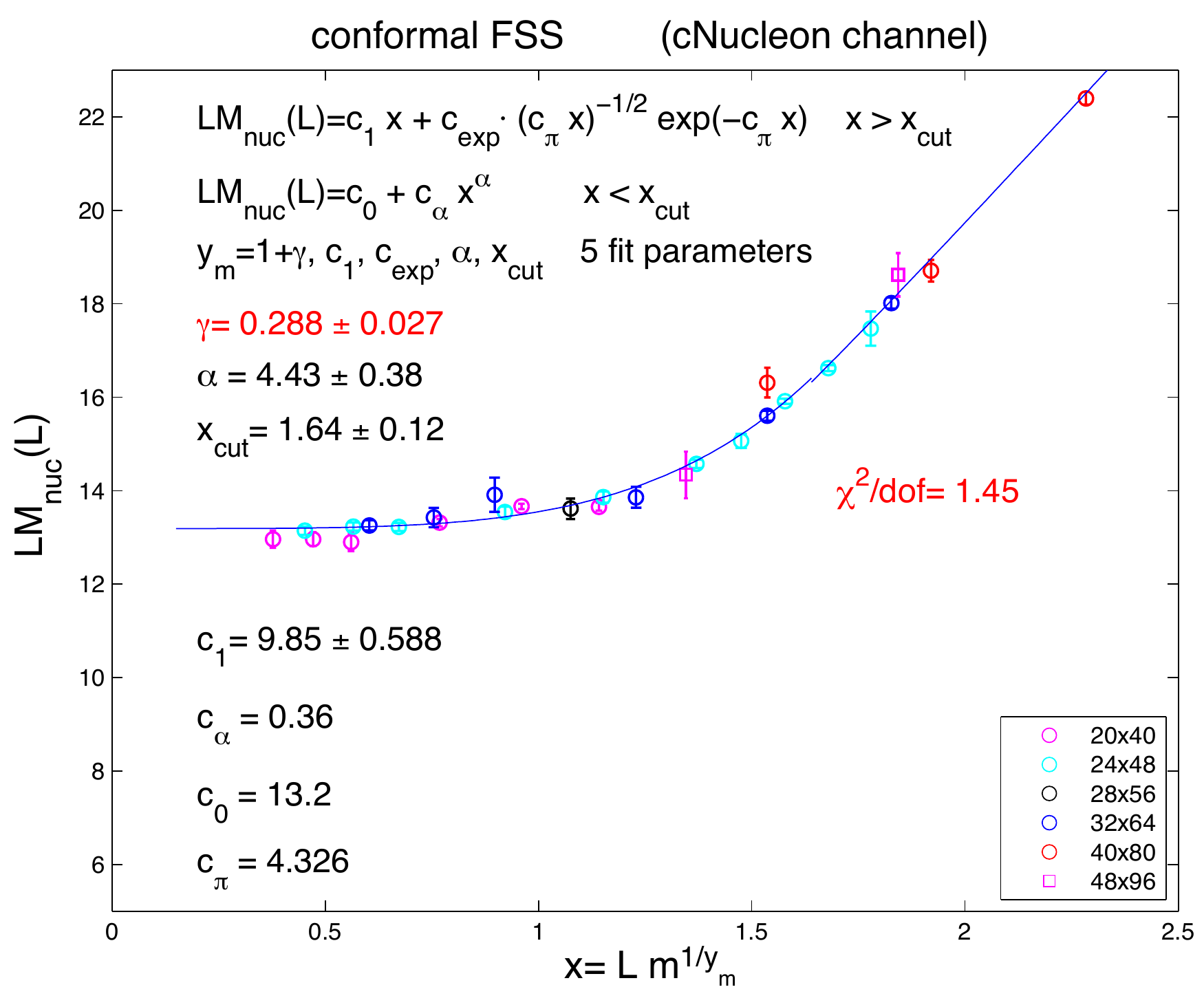}
\end{tabular}
\end{center}
\caption{{  Conformal FSS fits in four different quantum number channels. 
The fits are performed in each channel separately.
Since the $\gamma$ values vary considerably from channel to channel, a simultaneous global fit to the 
combined channels with the same $\gamma$ exponent, as required by conformal FSS theory, is bound to fail.}}
\label{fig:ConformFSS}
\end{figure}
The expected leading FSS  form for any mass $M$, or for $F_\pi$, scaled with the linear size $L$ of the spatial volume,
is given by a scaling function $L\cdot M=f(x)$ where $x=L\cdot m^{1/1+\gamma}$ is the conformal 
scaling variable. The scaling form sets in close to the critical surface for small $m$ values.
The scaling functions $f(x)$ can depend on the quantum numbers of the states 
but the scaling variable is expected to have the same form with identical $\gamma$ exponent in 
each quantum number channel~\cite{DelDebbio:2010hx,Bursa:2010xn,DelDebbio:2010ze,DelDebbio:2011kp}. 
In sub-leading order there are conformal FSS scaling violation effects which 
are exhibited as a combined cutoff and $L$-dependent leading  correction with the modified form 
$L\cdot M=f(x) + L^{-\omega}g(x)$ where the scaling correction exponent $\omega$ is determined at the 
infrared fixed point (IRFP) $g^*$
of the $\beta$-function as $\omega = \beta'(g^*)$. This assumes that the mass deformation away from the critical
surface is the only  relevant perturbation around the IRFP. The leading scaling correction term close enough 
to the critical surface dominates any other corrections which are supressed by further inverse powers of $L$.
To detect the leading scaling violation effect requires high precision data with fits to
scaling functions $f(x)$ and $g(x)$ and the critical exponent $\omega$. 

We applied conformal FSS theory to our data sets in the fermion mass range $m=0.006-0.035$ with lattice sizes ranging 
in the fits from  $20^3\times 40$ to $48^3\times 96$. Two different FSS fitting procedures were applied.
In the first procedure, we defined a scaling function $f(x)$ for each mass M with five independent
fitting parameters. The fitting function $f(x)$ was divided
into two regions separated at  the joint $x=x_{cut}$. Different forms were chosen on the two sides of $x_{cut}$ from 
the expected conformal behavior.
For large $x>x_{cut}$, the function $f(x)=c_1 x + c_{exp}(c_1x)^{-1/2}{\rm exp}(-c_1x)$ with
parameters $c_1$ and $c_{exp}$ describes the $L$-independent limit  $M \sim c_1 m^{1/1+\gamma}$ 
at fixed $m$ and $L\rightarrow\infty$.
The $c_{exp}$ amplitude sets the size of the leading small exponential correction from the 
finite volume effect of the lightest Goldstone pion state wrapping around the spatial volume.
Since $f(0)=c_0$ is expected from conformal FSS with some power corrections at small $x$, we applied the simple
ansatz $f(x)=c_0 + c_{\alpha}x^{\alpha}$ for $x<x_{cut}$ (a more general polynomial function in the small $x$ region is not
expected to change the conclusions from the fits).
From the fit to the PCAC Goldstone pion channel the parameter $c_{\pi}=c_1$ was determined and used as input in 
the exponential terms  of the other channels with ${\rm exp(-c_\pi L}$).
The critical exponent $\gamma$ was included among the five fitting parameters, in addition to $c_0,~c_1,~c_{exp}$, and $x_{cut}$.

The composite particle masses in several  quantum number channels can be reasonably fitted 
with  conformal scaling functions $f(x)$ as shown in Figure~\ref{fig:ConformFSS} but the values of the 
critical exponent $\gamma$ are incompatible across different channels. The required global conformal FSS fit will fail
with a single exponent $\gamma$ across all quantum numbers.
In the fits for $F_\pi$ in the PCAC pion channel we only kept four parameters because the asymptotic form with exponentially 
small correction was zero within error. Actually, the data of $F_\pi$ did not allow a successful conformal fit with any shape chosen
for its scaling function $f(x)$ which looks very different from the scaling functions of composite particle masses. The unexpectedly
curious behavior of the $F_{\pi}$ data set against conformal FSS remains unresolved.

\subsection{Generalized FSS fitting procedure with spline based general B-form}

Following a new  fitting strategy, we investigated if the failed global conformal FSS analysis can 
be attributed to restrictions on the conformal 
scaling functions $f(x)$. The restrictions were  manifest in the physics-motivated fitting procedure we applied above. 
Our new general approach is different from~\cite{Appelquist:2011dp,DeGrand:2011cu} but addresses related issues. 
We developed a general least-squares fitting
procedure to the scaling functions using the B-form of spline functions without any further restrictions. 
In this procedure, the function $f(x)$ is described
by piece-wise polynomial forms constructed from spline base functions with general
coefficients in overlapping intervals of the scaling variable $x$. 
The shape of the B-form can be changed without limitations by increasing the number of base functions and the number
of scaling intervals in $x$  bracketing the overlapping data range.
The details of this new analysis will be reported elsewehere~\cite{Fodor:prep}.  
\begin{figure}[h]
\begin{center}
\begin{tabular}{ccc}
\includegraphics[height=4.1cm]{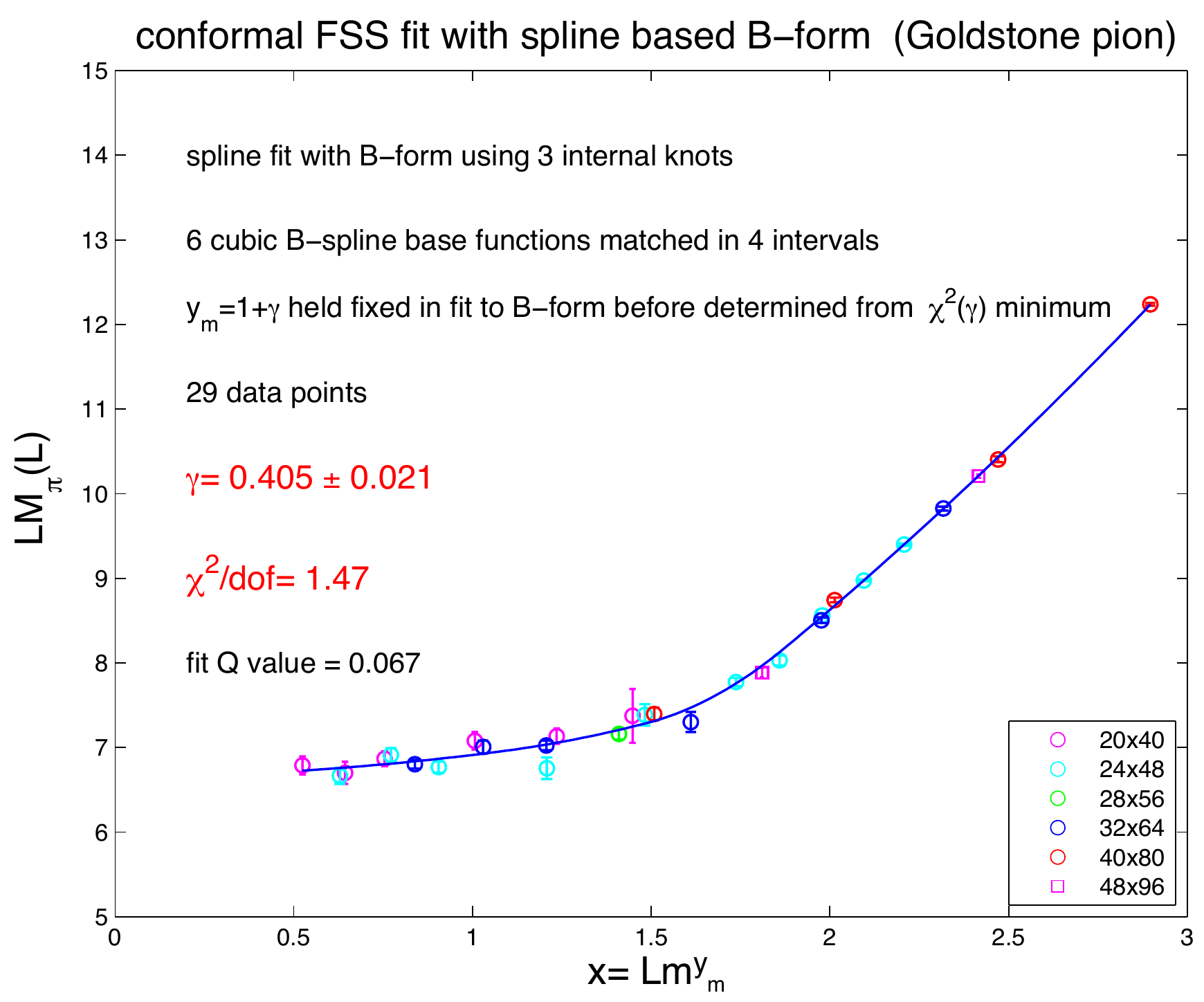}&
\includegraphics[height=4.1cm]{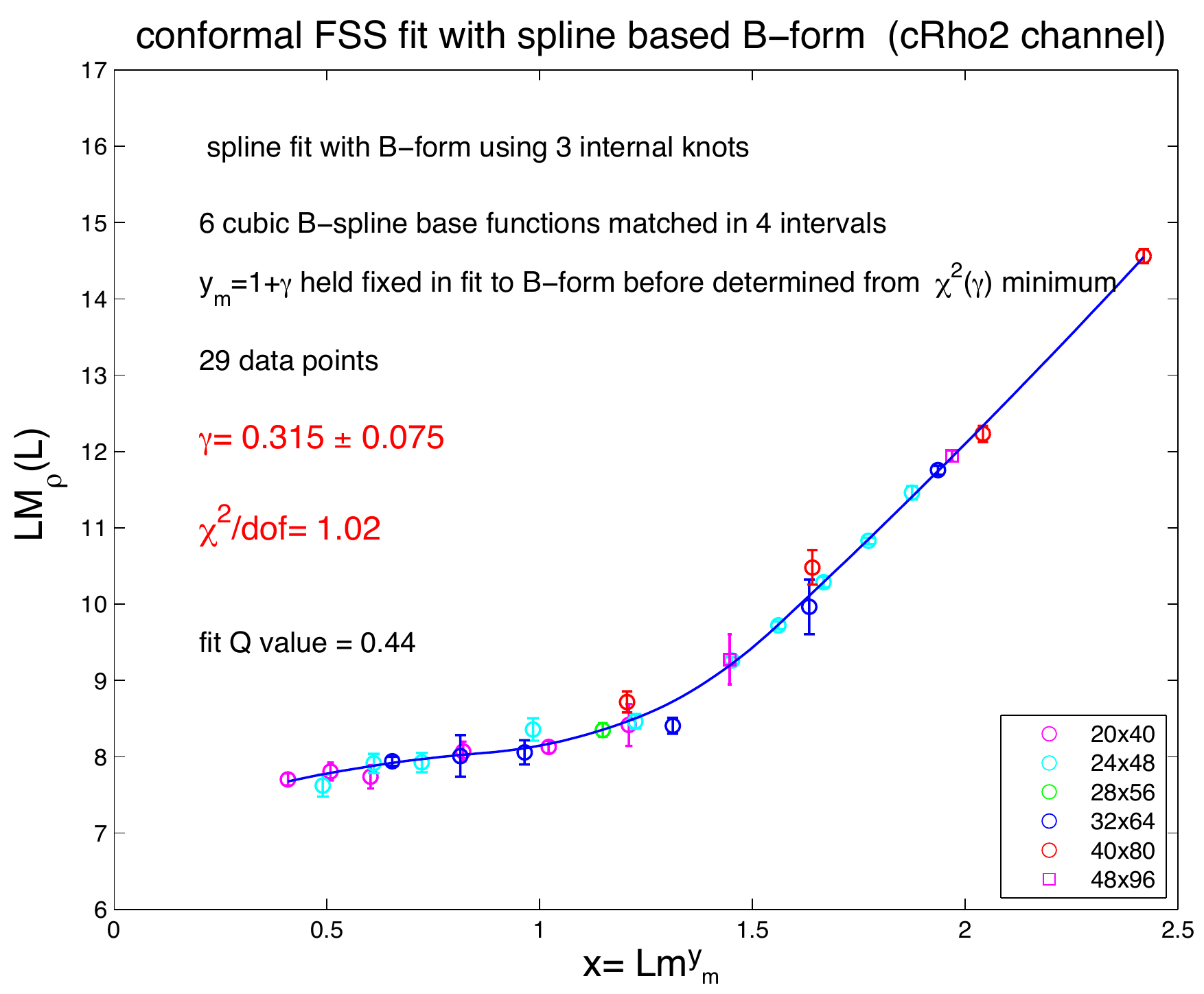}
\includegraphics[height=4.1cm]{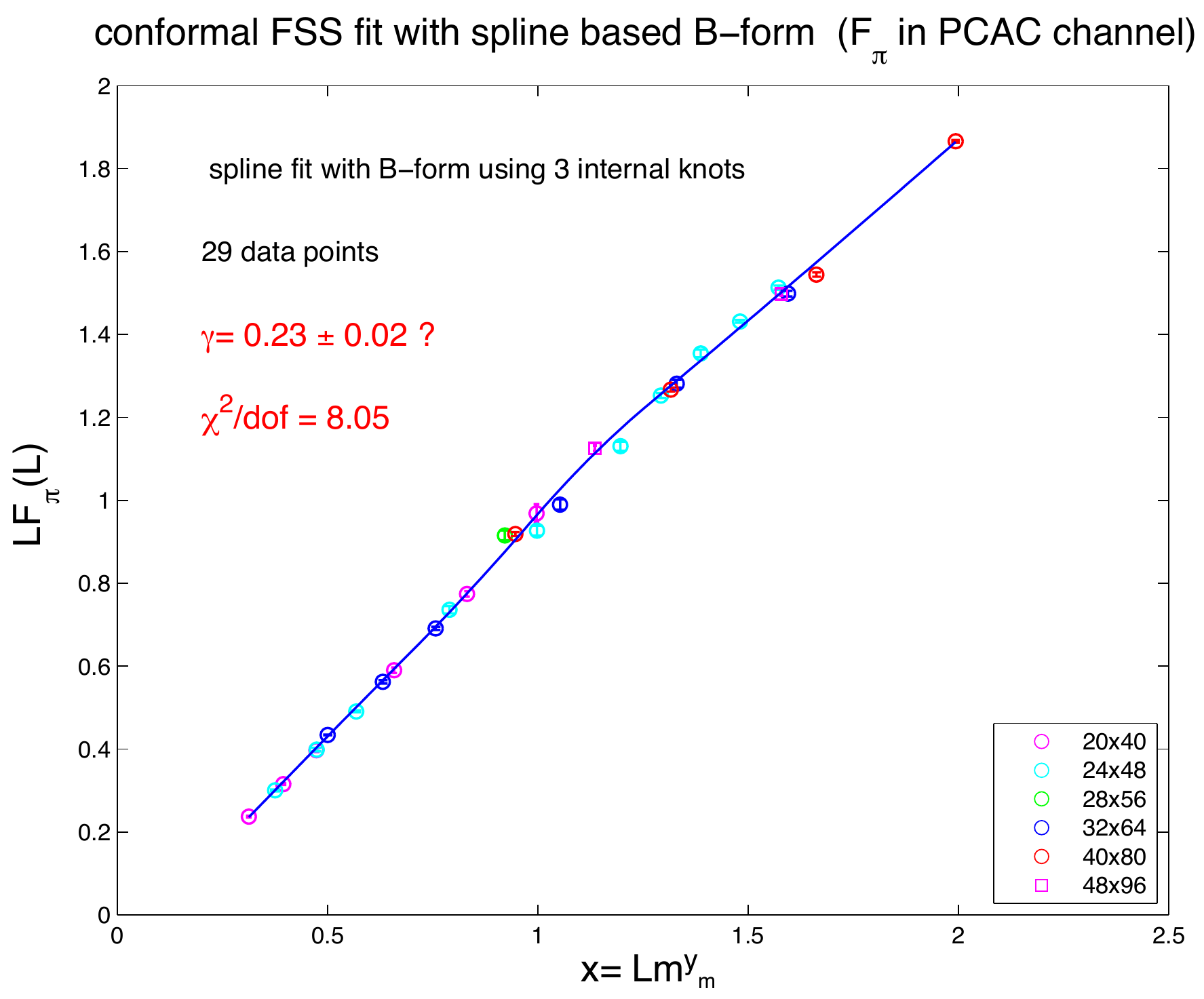}&
\end{tabular}
\end{center}
\caption{{\footnotesize  Conformal FSS fits using spline based B-forms in three different channels. 
The fits are preformed in each channel separately with the question mark on $\gamma$ indicating
difficulties of error estimates in bad fits of  $F_\pi$ . }}
\label{fig:splineConformFSS}
\end{figure}

Our fitting procedure in its setup requires two steps.
In the first step, for any fixed choice of the exponent $\gamma$, the 
best fitted function $f(x)$ is determined in spline function B-form from the minimization of the weighted $\chi^2$ expression.
According to a general algorithm, the $x$-range of the data set is divided into intervals separated by internal knots and 
adding end point knots for B-form spline construction.  The number of coefficients is determined by the number of knots and the order of 
the spline polynomials of the sub-intervals. 
The weighted $\chi^2$ sum is minimized with respect to the coefficients of the base functions in the B-form. This will produce
the best fit for fixed $\gamma$ with a minimized $\chi^2$ sum which will depend on $\gamma$.
In the second step, we minimize the $\chi^2$ sum with respect to $\gamma$ to determine the best fit
of the critical exponent. The one-$\sigma$ confidence interval is determined from the variation of the $\chi^2$ sum as a function
of $\gamma$.

In Figure~\ref{fig:splineConformFSS} we show three typical fits for illustration. The fit to the Goldstone pion in the PCAC channel 
improved as expected, with considerable increase in the error. The tension across channels decreased, as illustrated by
comparison with the rho-channel fit, but the fit to $F_{\pi}$ remained unacceptable. 
With the extended data set we are unable to reproduce results in ~\cite{Appelquist:2011dp,DeGrand:2011cu}
which used tables from our earlier limited subset of data~\cite{Fodor:2011tu} in favor of consistency with the conformal phase.
It is important to emphasize that we have not reached 
definitive conclusions about the failure of conformal tests.  As we stated earlier~\cite{Fodor:2011tu}, 
we have not analyzed yet the leading scaling violation effects and did not investigate if the good scaling form in separate 
quantum number channels can be explained in the chirally broken phase by strongly sqeezed wave function effects. 
In disagreement with ~\cite{DeGrand:2011cu}, conformal FSS based analysis of the spectrum and related 
sum rules on moments of the correlators we have been developing are deep renormalization group based probes 
of the conformal phase. 
 As explained in our forthcoming publication~\cite{Fodor:prep}, we remain skeptical about the fitting procedure followed in
~\cite{Appelquist:2011dp} with efforts to rescue the conformal interpretation. The issues are not settled and ultimately 
will be decided in more definitive analyses.

\section{Two fermion flavors in the sextet  SU(3) color representation}

This model has been studied recently by three  BSM groups~\cite{DeGrand:2010na,DeGrand:2012yq,Sinclair:2010be,Fodor:2011tw}.
Our findings are different from results based on  the Schr\"odinger functional~\cite{DeGrand:2010na,DeGrand:2012yq} and
compatible with the finite temperature
phase transition in~\cite{Sinclair:2010be}. The disagreement with  Schr\"odinger functional results
is particularly significant based on the lower bound  $\gamma \geq 1$ we find adopting the conformal hypothesis.
This can be important in BSM applications and remains in contrast with the small exponent $\gamma < 0.45$ 
published in~\cite{DeGrand:2012yq}.

We have new simulation results at $\beta=3.2$ in the fermion mass range ${\rm m=0.003-0.010}$ on
$24^3\times48$ and $32^3\times64$ lattices. Five fermion masses
at  ${\rm m=0.003,0.004,0.005,0.006,0.008}$ are used in most fits.
For further checks on finite volume dependence, a very large and expensive $48^3\times96$ 
run was added recently at ${\rm m=0.003}$ to follow the strategy of finite volume extrapolation
at fixed fermion mass $m$.
We also have new preliminary simulation results at $\beta=3.25$ in the mass range ${\rm m=0.004-0.008}$ on
$24^3\times48$ and $32^3\times64$ lattices. Based on the chiral and conformal analyses of the model, continued runs
at existing run parameters and new runs are planned at both couplings to further probe the conformal
FSS hypothesis in the sextet model following the strategy we presented for the $N_f=12$ model.

\subsection{Finite volume extrapolation}
\begin{figure}[htb]
\begin{center}
\begin{tabular}{ccc}
\includegraphics[height=3.9cm]{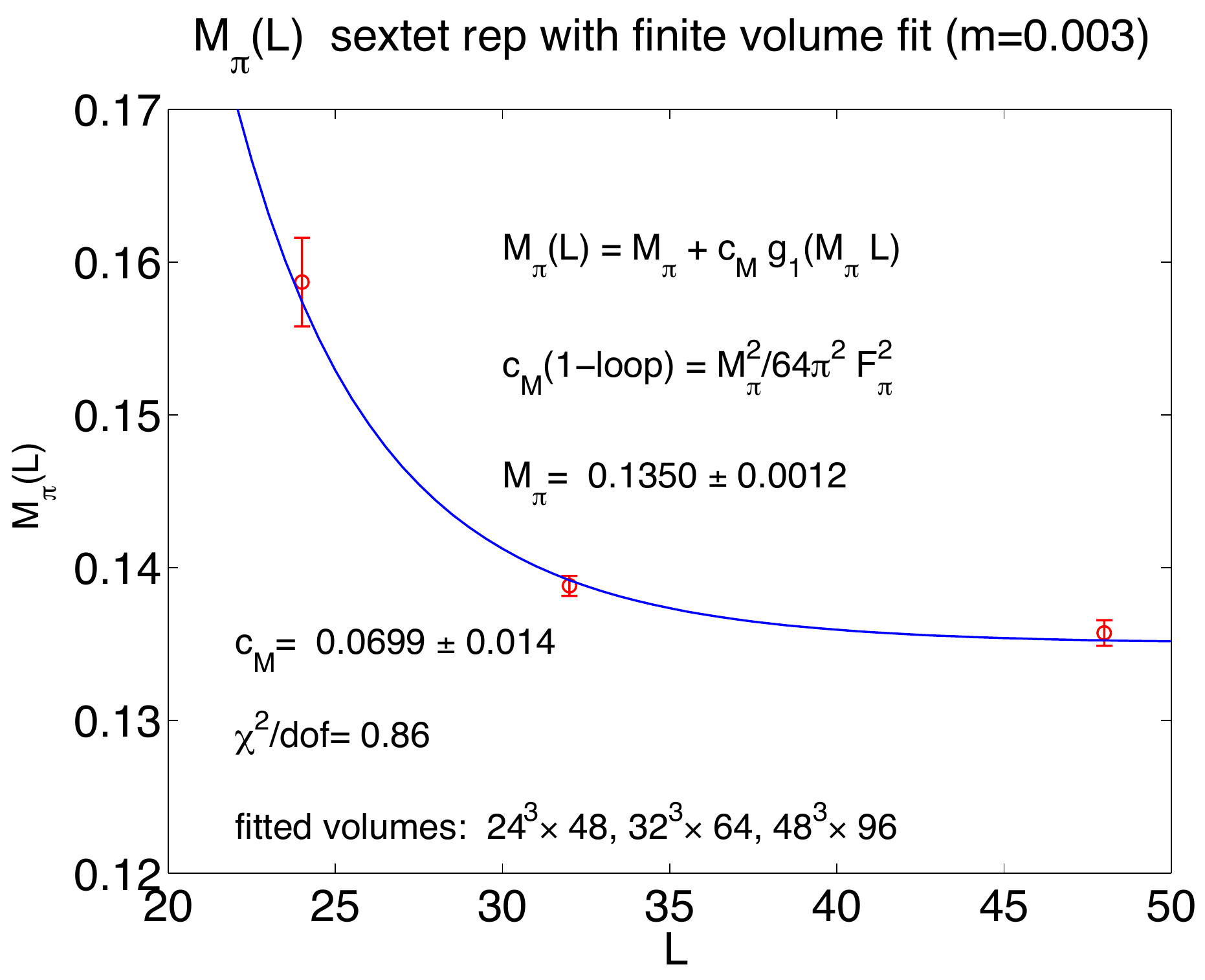}&
\includegraphics[height=3.9cm]{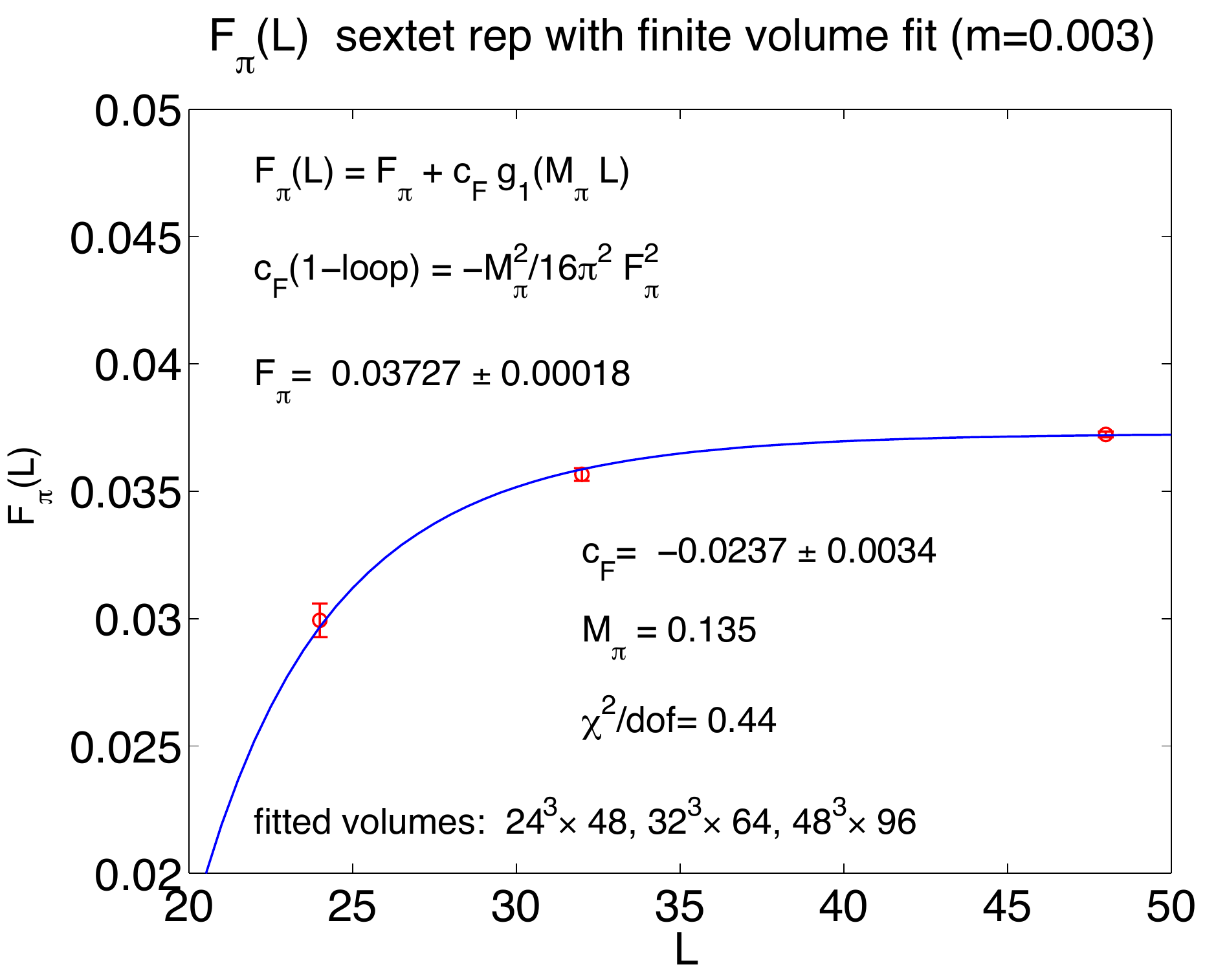}&
\includegraphics[height=3.9cm]{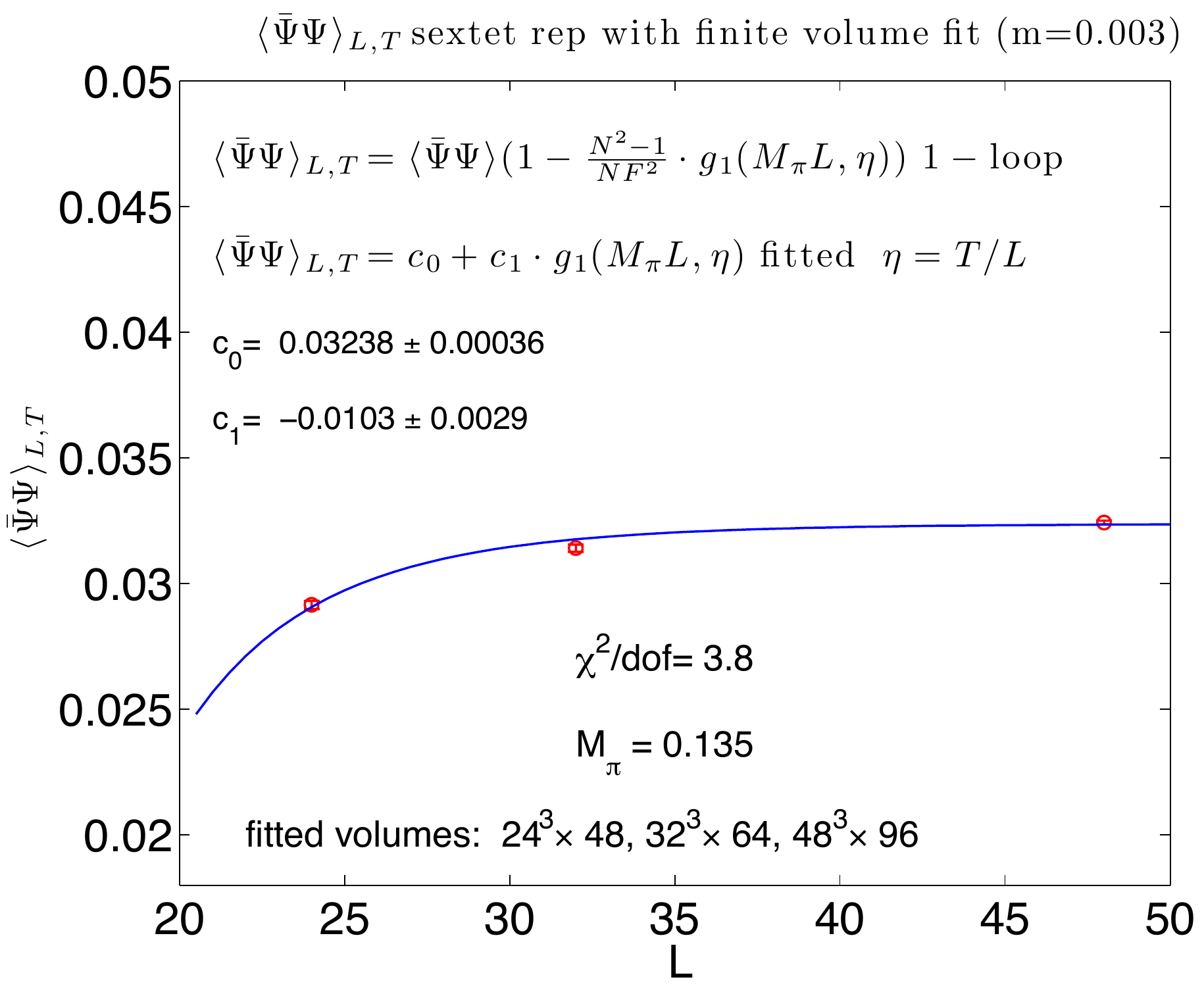}
\end{tabular}
\end{center}
\caption{\footnotesize  Finite volume dependence at the lowest fermion mass for $\beta=3.2$.
The form of $\tilde g_1(\lambda,\eta)$ is a complicated infinite sum which contains Bessel
functions and requires numerical evaluation~\cite{Gasser:1986vb}. Since we are not in the chiral log regime, the prefactor of
the $\tilde g_1(\lambda,\eta)$ function was replaced by a fitted coefficient. The leading term of  the function
$\tilde g_1(\lambda,\eta)$ is a special exponential Bessel function $K_1(\lambda)$ which dominates in the simulation range.}
\label{fig:sextetInfVol}
\end{figure}
Based on the $\chi{\rm SB}$ hypothesis,  infinite volume extrapolations of the Goldstone pion, $F_\pi$, and $\langle\bar\psi\psi\rangle$
are shown in Figure~\ref{fig:sextetInfVol} where $\tilde g_1(\lambda,\eta)$ describes the finite volume corrections with
$\lambda=M_\pi\cdot L$ and aspect ratio $\eta=T/L$ from the lightest pion~\cite{Leutwyler:1987ak}.  
The fitting procedure approximates the leading treatment of  the pion which wraps around the finite volume,
whether in chiral perturbation theory, or in L\"uscher's non-perturbative finite volume analysis~\cite{Luscher:1985dn}. 
This does not require to reach the 1-loop  chiral log limit
as long as the pion is the lightest state dominating the finite volume corrections.
The infinite volume limits of $M_\pi$, $F_\pi$,  and $\langle\bar\psi\psi\rangle$ for  $m=0.003$ at $\beta=3.2$
were determined self-consistently from the fitting procedure. Similar fits were applied to other composite states.
Based on the fits at $m=0.003$,  one percent accuracy of the infinite volume limit is reached at $M_\pi L= 5$.
In the fermion mass range $m \geq 0.004$ the condition $M_\pi L> 5$ is reached at $L=32$.
Although it will require high precision runs to test, we do not expect more than one percent residual volume 
dependence in the  $32^3\times64$ runs for $m \geq 0.004$.
Based on these observations, we will present chiral and conformal analyses with 
extrapolated infinite volume scaling behavior from the $32^3\times64$ runs for $m \geq 0.004$.

\subsection{The chiral condensate and $\chi SB$} 
We follow the analysis of the chiral condensate as described for the $N_f=12$ model.
The $\langle \bar{\psi}\psi\rangle$ condensate data were fitted with a third order 
polynomial of the form $c_0+c_1m+c_2m^3$ while the condensate with derivative subtraction
was fitted without the linear term. Both independently measured quantities have to converge  
to the same chiral limit.
\begin{figure}[hbt]
\begin{center}
\begin{tabular}{ccc}
\includegraphics[height=5.5cm]{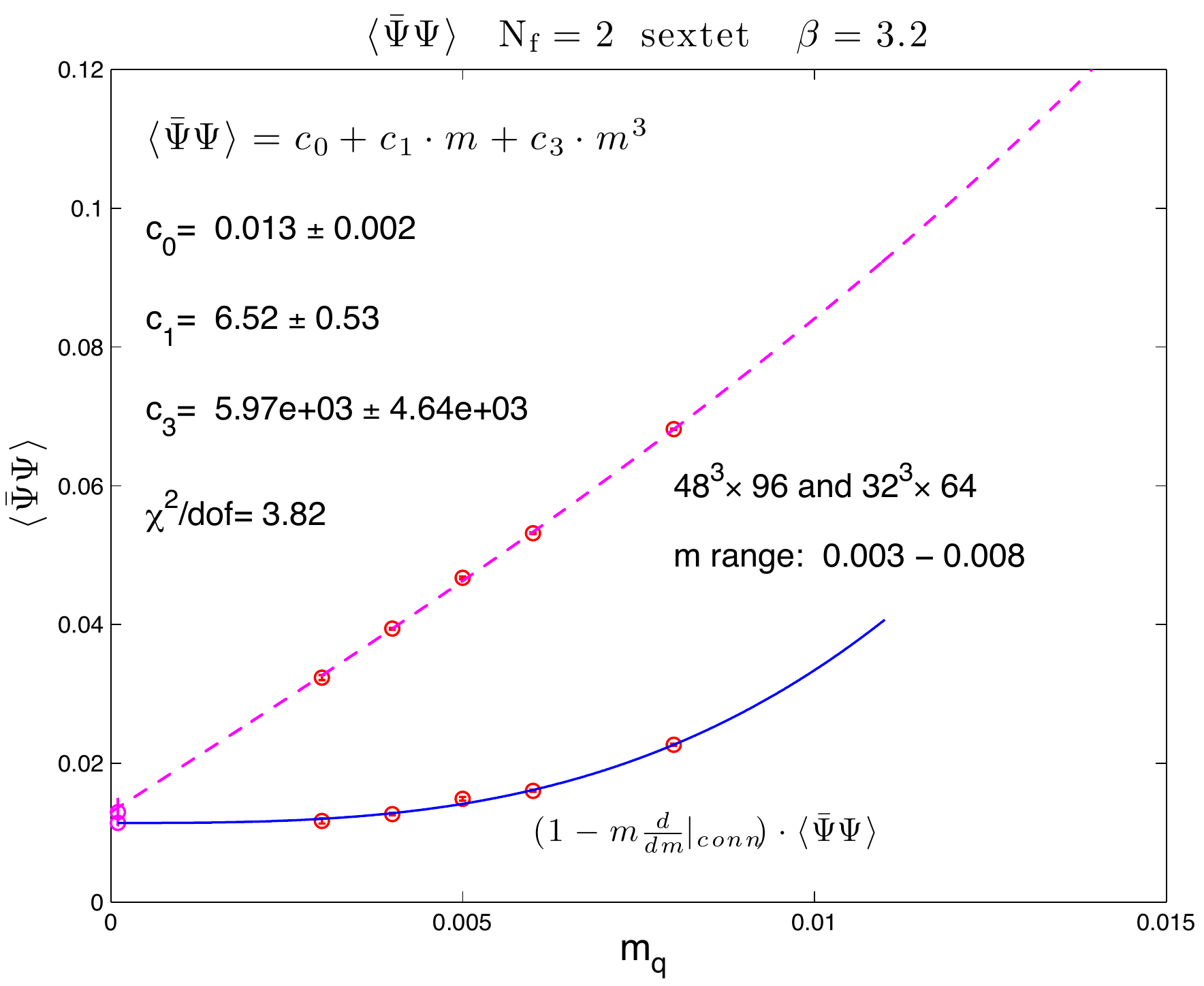}&
\includegraphics[height=5.5cm]{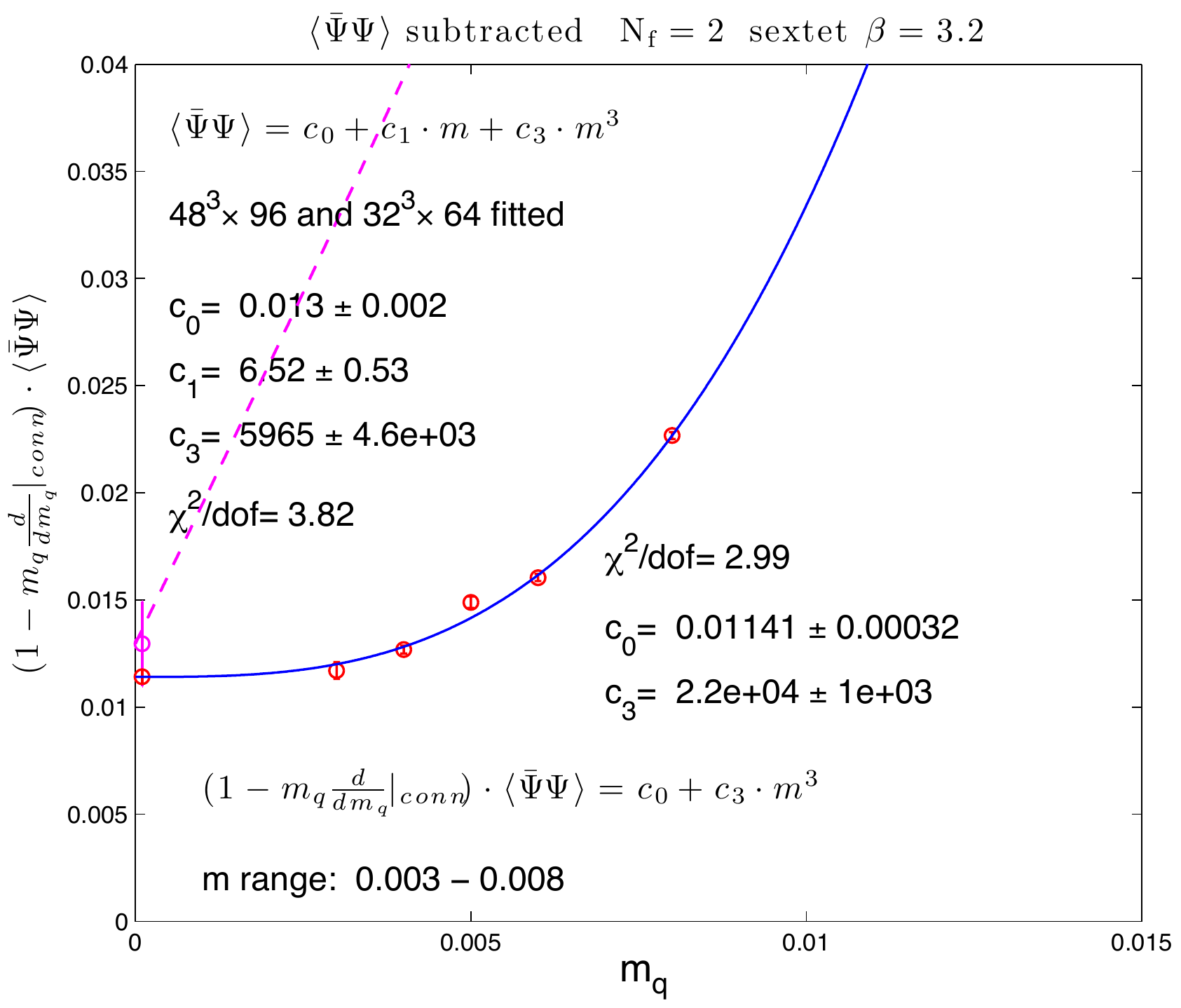}&
\end{tabular}
\end{center}
\vskip -0.1in
\caption{\footnotesize
For any given $m\geq 0.004$  the largest volume condensate data is used since 
the finite volume analysis remains incomplete. The two plots are discussed in the text.}
\label{fig:PbPsextet}
\end{figure}
The chiral condensate and its subtracted derivative version  are shown in Figure~\ref{fig:PbPsextet}
with a consistent strong $\chi SB$ signal in the chiral limit. 

\subsection{Spectrum and the ${\bf\rm\chi SB}$ hypothesis}
\begin{figure}[h]
\begin{center}
\begin{tabular}{cc}
\includegraphics[height=5cm]{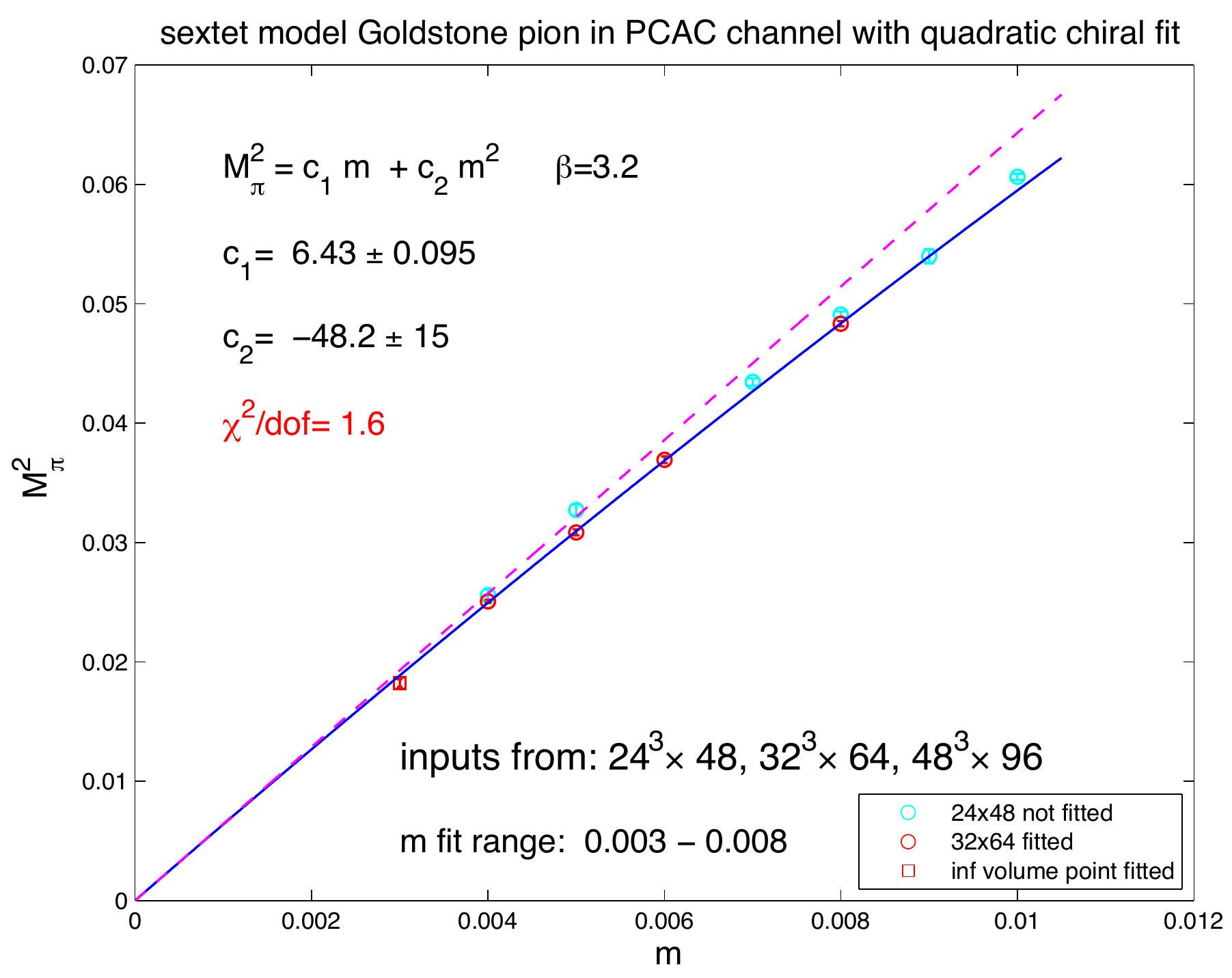}&
\includegraphics[height=5cm]{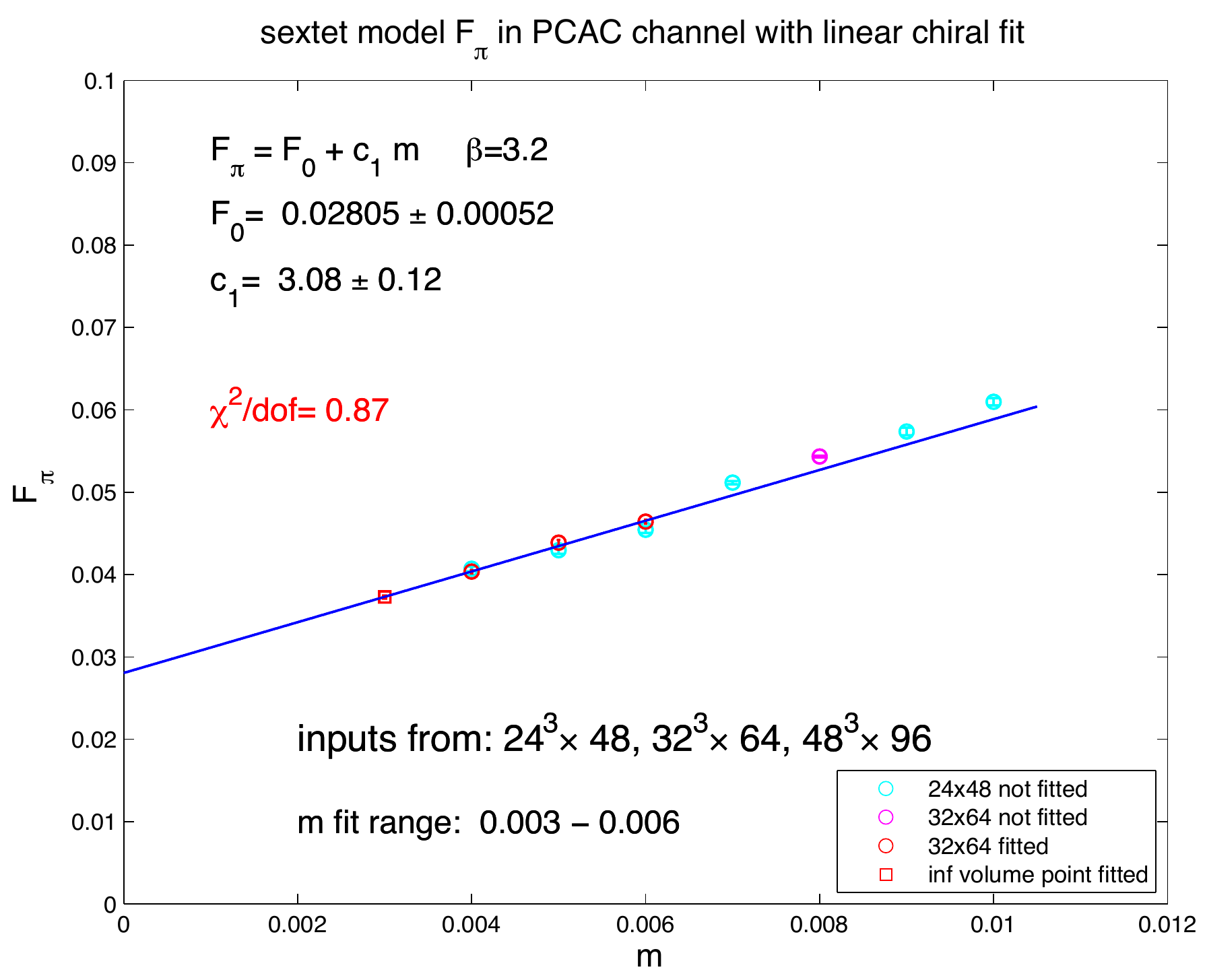}\\
\includegraphics[height=5cm]{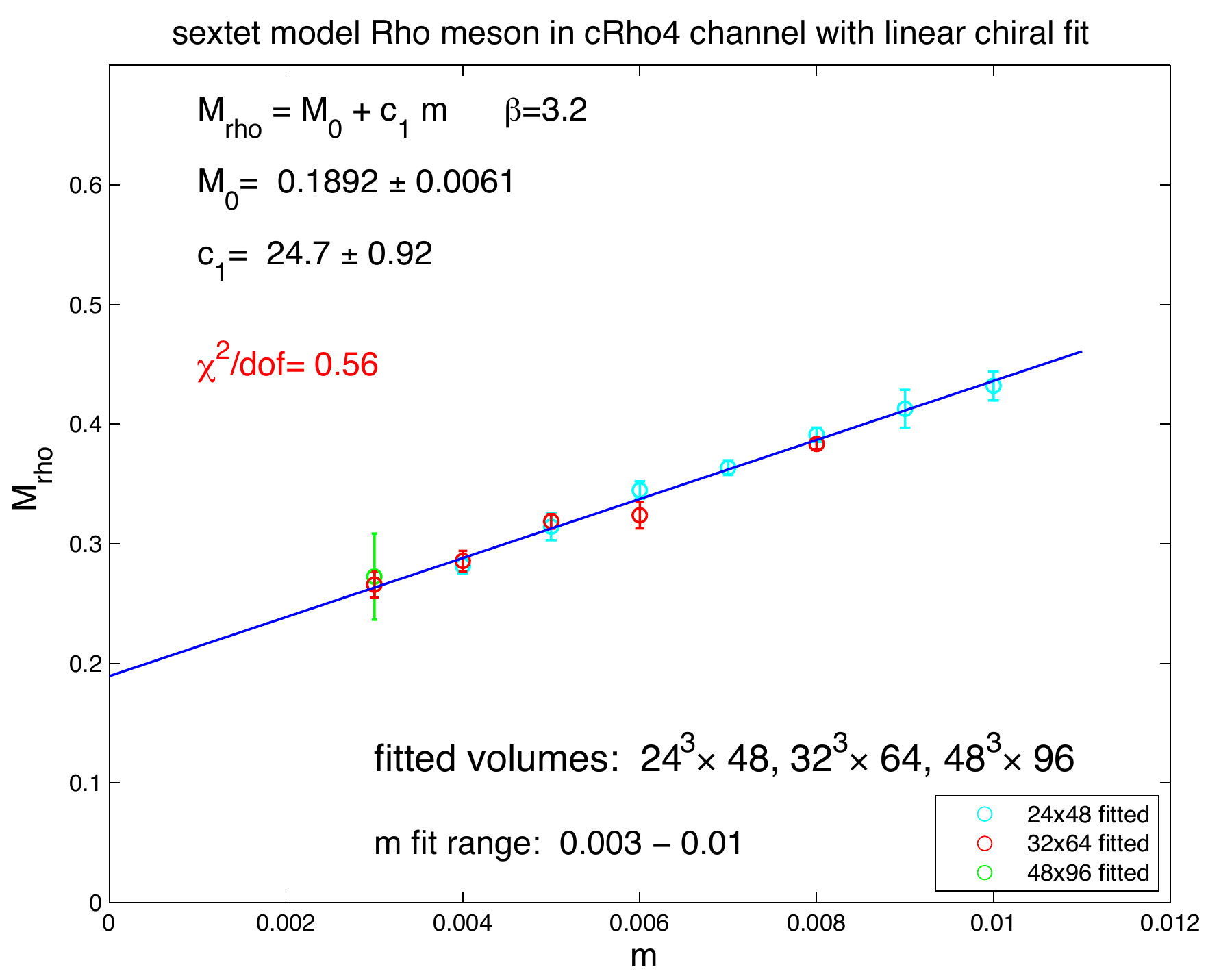}&
\includegraphics[height=5cm]{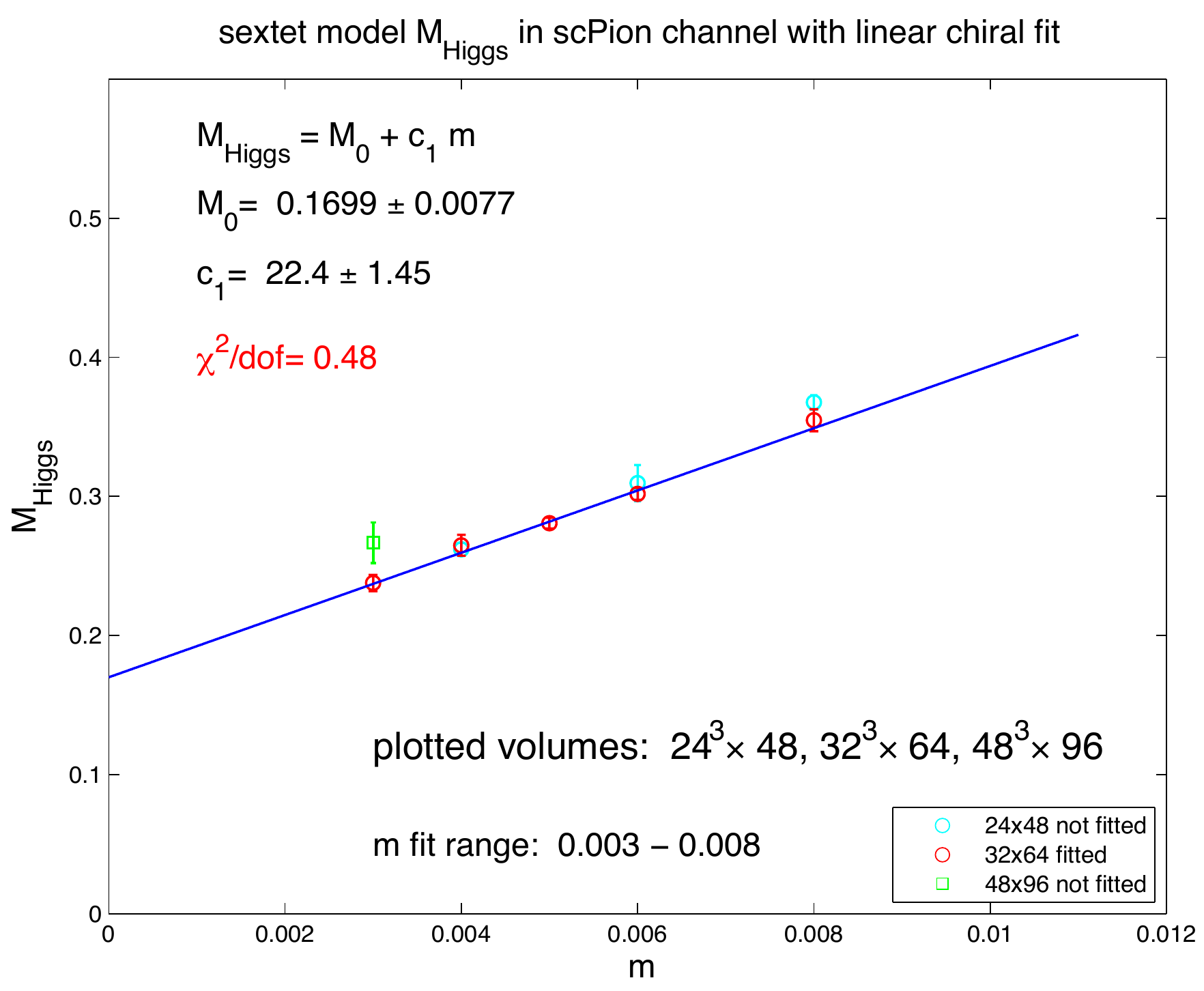}
\end{tabular}
\end{center}
\caption{\footnotesize  Polynomial fits from the analytic mass dependence of the chiral Lagrangian without logarithmic 
loop corrections are shown for the Goldstone pion, $F_\pi$,  $M_\rho$, and the $0^{++}$ state with mass  $M_{Higgs}$. 
The dashed line in the Goldstone pion
plot shows the leading linear contribution.
$F_0$ on the top right plot sets the eletroweak vev scale. The disconnected diagram, 
which can shift the final value of the Higgs mass  from  $M_0= 6.06\cdot F_0$ on the bottom right plot, is not included in
the calculation.}
\label{fig:sextetChiralSpectrum}
\end{figure}
Based on the analytic fermion mass dependence of the chiral Lagrangian, 
and using the lowest fermion masses in the $m=0.003-0.008$ range,
good polynomial fits were obtained without logarithmic loop corrections as shown 
in Figure~\ref{fig:sextetChiralSpectrum} for four select states. 
The plotted $24^3\times 48$ data points for $m \geq 0.004$ agree with the fitted data from the $32^3\times 64$ runs
indicating the infinite volume limit within the accuracy of the data. Small corrections, if  required, 
should not effect the conclusions. 
Although we could fit $M_\pi$ and $F_\pi$ with the continuum chiral
logarithms included, the separate sets of $F$ and $B$ from the fits are not quite self-consistent. A combined staggered SU(2)
chiral perturbation theory fit is required for simultaneous fits of $M_\pi$ and $F_\pi$ with a consistent pair of fundamental
chiral parameters $F$ and $B$.
The explicit cutoff dependent corrections to the $F$ and $B$ parameters would require further testing at weaker gauge couplings
and using partially quenched staggered chiral perturbation theory. Our runs at $\beta=3.25$ should provide the data for this analysis.

\subsection{Conformal hypothesis and the critical exponent $\bf\gamma$}
It is important to compare the polynomial fits with conformal scaling behavior for small mass deformations $m$. 
In the infinite volume limit the masses of composite particles and $F_\pi$ are expected to scale 
as $ M \sim m^{1/1+\gamma}$ with the same exponent $\gamma$ in all channels.
\begin{figure}[h]
\begin{center}
\begin{tabular}{ccc}
\includegraphics[height=3.9cm]{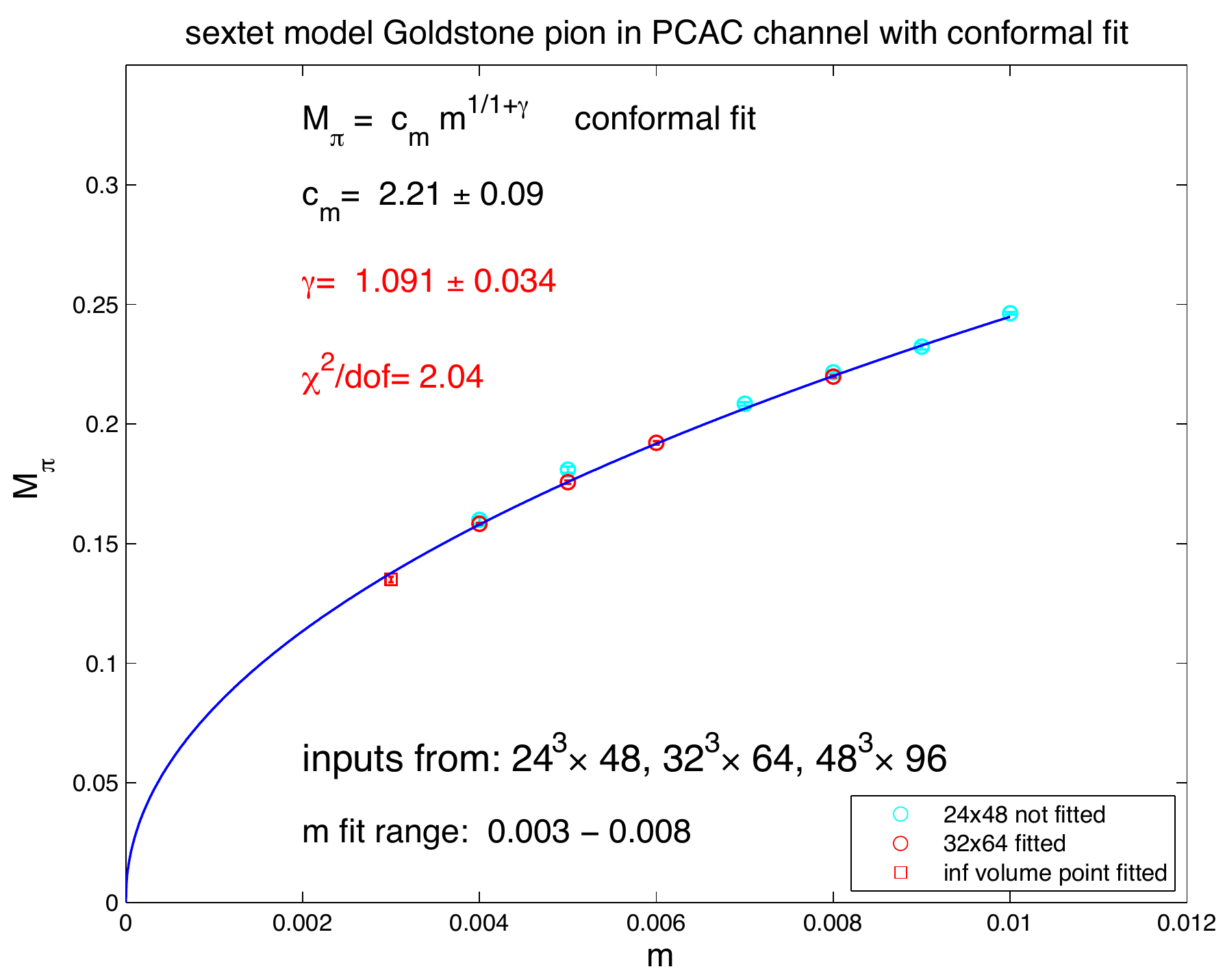}&
\includegraphics[height=3.9cm]{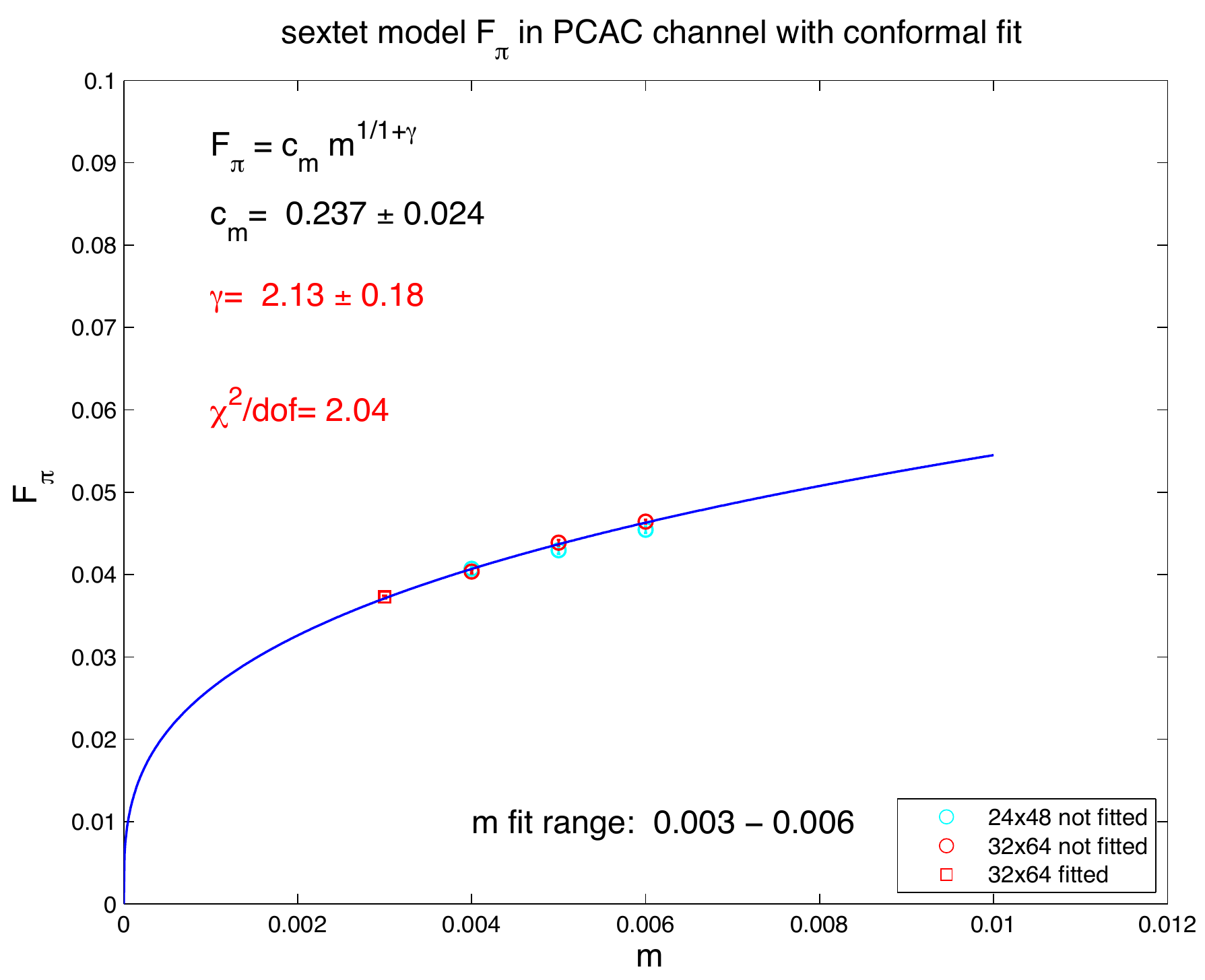}&
\includegraphics[height=3.9cm]{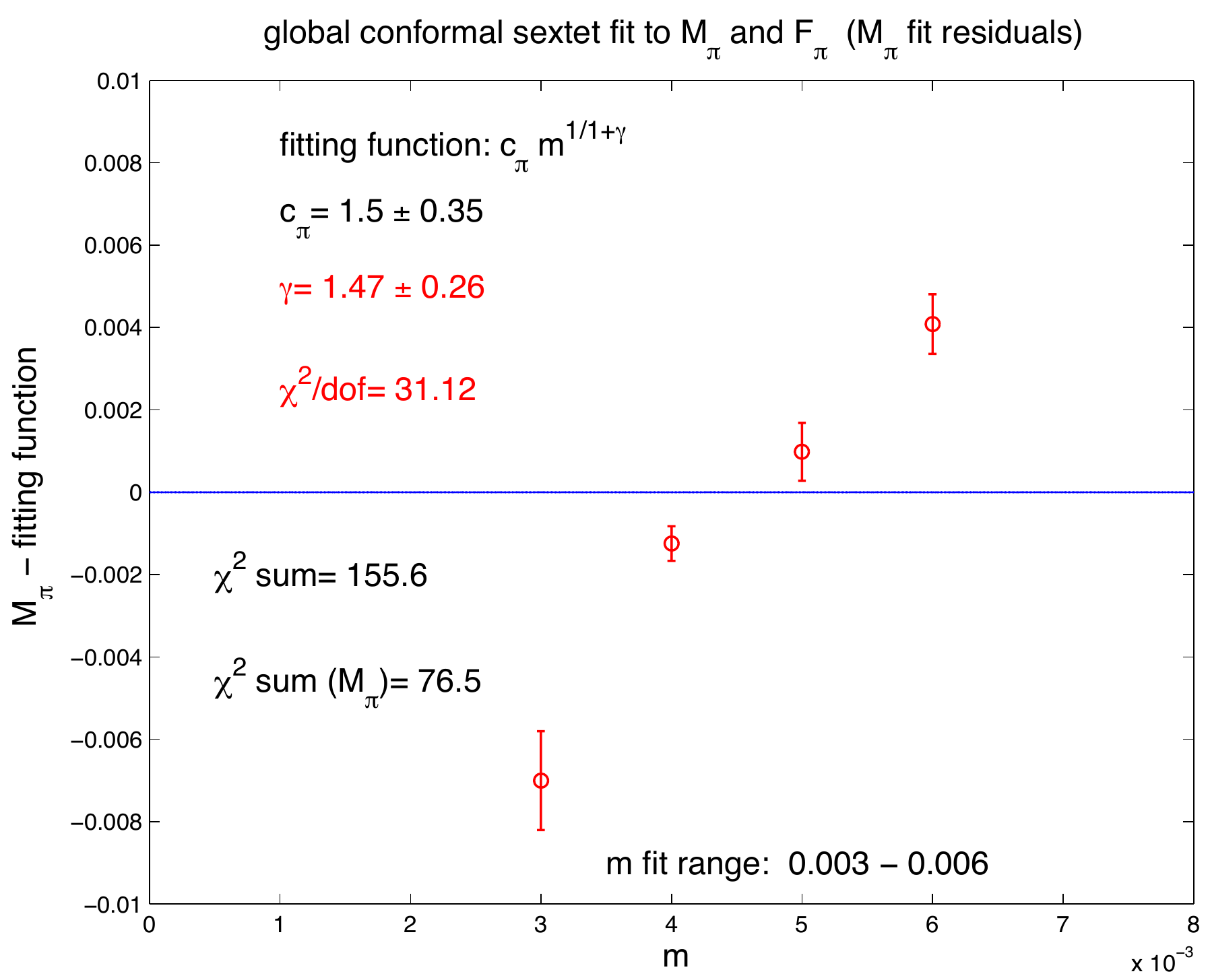}
\end{tabular}
\end{center}
\caption{\footnotesize The left side plot and the middle plot represent separate conformal fits. 
The right side plot display the $M_\pi$ residuals from the global fit. It is unacceptable for $F_\pi$ as well.
The global fit is trying to choose a $\gamma$ value between $\gamma \sim 1$ in the Goldstone pion
channel and $\gamma \sim 2$ in the $F_\pi$ fit resulting in a very high $\chi^2$ value. All fits are at $\beta=3.2$.
}
\label{fig:sextetConformTest}
\end{figure}
%
When the four lowest fermion mass values closest to the critical surface are fitted separately 
with the leading conformal form, we get good $\chi^2$ fits but very different $\gamma$ exponents, which is not consistent
with mass deformed conformal behavior.
The conflicting fits are illustrated side by side in Figure~\ref{fig:sextetConformTest} for the 
Goldstone pion and the $F_\pi$ decay constant. Fitting to the pion mass requires $\gamma=1.091(34)$ while the 
$F_\pi$ fit is forcing  $\gamma=2.13(18)$. In the combined fit they compromise with  $\gamma=1.47(26)$
and the unacceptable ${\rm \chi^2/dof=31.1}$.
It is very difficult to see how this conflict, also in disagreement with ~\cite{DeGrand:2012yq}, 
could be resolved within the conformal hypothesis. 
From the tests we were able to perform, the sextet model is consistent with $\chi SB$ and inconsistent with conformal symmetry. 
It will require further investigations to show that subleading effects cannot alter this conclusion. 
We will consider comprehensive conformal FSS tests which do not rely on 
infinite volume extrapolation in the scaling fits. This is at a preliminary stage requiring new runs and 
systematic analysis. 

If $\chi SB$ of the sextet model is further confirmed in the massless fermion limit, its relevance for the
realization of the composite Higgs mechanism is transparent. 
The large anomalous exponent $\gamma$ of our conformal fits will be interpreted in this case as an important
ingredient of the model in the $\chi SB$  phase. Importantly,
the model has the perfect match 
of three Goldstone pions to provide the longitudinal components of the $W^{\pm}$ and $Z$ bosons.
To understand the slowly changing
gauge coupling close to the conformal window without infrared fixed point will 
require high precision methods to calculate the renormalized gauge coupling and its beta function.
This will demand extended and more reliable Schr\"odinger functional analysis or alternate methods 
which are being developed. The difference between the large
exponent $\gamma$ reported here and the low value of $\gamma$  published earlier~\cite{DeGrand:2012yq}
is significant and will require clarifications. Conformal FSS tests very close to the critical surface will
provide further independent checks of our results.

\acknowledgments{We are grateful to Kalman Szabo and Sandor Katz
for their code development and to Anna Hasenfratz for discussions. 
The simulations were performed using computational resources
at Fermilab and JLab, under the auspices of USQCD and SciDAC. 
This research is supported by
the NSF under grants 0704171 and 0970137, by the DOE under grants
DOE-FG03-97ER40546, DOE-FG-02-97ER25308, by the DFG under 
grant FO 502/1 and by SFB-TR/55, and the EU Framework Programme 7 grant
(FP7/2007-2013)/ERC No 208740.
Some of the simulations used allocations from 
the Extreme Science and Engineering Discovery Environment (XSEDE), 
which is supported by National Science Foundation grant number OCI-1053575.


\end{document}